\documentclass[times,5p,twocolumn,final,authoryear]{elsarticle}

\usepackage{medima}
\usepackage{framed,multirow}

\usepackage{amssymb}
\usepackage{latexsym}

\usepackage{xcolor}

\usepackage{hyperref}

\definecolor{newcolor}{rgb}{.8,.349,.1}

\usepackage{multirow}
\usepackage{ulem}

\usepackage{graphicx}
\usepackage{bm}
\usepackage{threeparttable}

\journal{Medical Image Analysis}

\begin{document}

\verso{L. Mou \textit{et~al.}}

\begin{frontmatter}

\title{CS$^2$-Net: Deep Learning Segmentation of Curvilinear Structures in Medical Imaging}%

\author[1]{Lei \snm{Mou}}
\author[1]{Yitian \snm{Zhao}*}
\ead{yitian.zhao@nimte.ac.cn}
\author[3]{Huazhu \snm{Fu}}
\author[4]{Yonghuai \snm{Liu}}
\author[5]{Jun \snm{Cheng}}
\author[6,1]{Yalin \snm{Zheng}}
\author[1]{Pan \snm{Su}}
\author[1]{Jianlong \snm{Yang}}
\author[7]{Li \snm{Chen}} 
\author[1,10,8]{Alejandro  F \snm{Frangi}}
\author[11]{Masahiro  \snm{Akiba}}
\author[2,1,9]{Jiang \snm{Liu}*}
\ead{liuj@sustech.edu.cn}

\address[1]{Cixi Institute of Biomedical Engineering, Ningbo Institute of Materials Technology and Engineering, Chinese Academy of  Sciences, Ningbo, China}
\address[2]{Department of Computer Science and Engineering, Southern University of Science and Technology, Shenzhen, China}
\address[3]{Inception Institute of Artificial Intelligence, Abu Dhabi, United Arab Emirates}
\address[4]{Department of Computer Science, Edge Hill University, Ormskirk, UK}
\address[5]{UBTech Research, UBTech Robotics Corp Ltd, Shenzhen, China}
\address[6]{Department of Eye and Vision Science, University of Liverpool, Liverpool, UK}
\address[7]{School of Computer Science and Technology, Wuhan University of Science and Technology, Wuhan, China}
\address[8]{Centre for Computational Imaging and Simulation Technologies in Biomedicine (CISTIB), School of Computing and School of Medicine, University of Leeds, Leeds, UK; Leeds Institute of Cardiovascular and Metabolic Medicine, School of Medicine, University of Leeds, Leeds, UK}
\address[9]{Guangdong Provincial Key Laboratory of Brain-inspired Intelligent Computation, Department of Computer Science and Engineering, Southern University of Science and Technology, Shenzhen, China}
\address[10]{Medical Imaging Research Centre (MIRC), University Hospital Gasthuisberg. Cardiovascular Sciences and Electrical Engineering Departments, KU Leuven, Leuven, Belgium}
\address[11]{R\&D Division, Topcon Corporation, Japan}



\begin{abstract}
Automated detection of curvilinear  structures, e.g., blood vessels or nerve fibres, from medical and biomedical images is a crucial early step in automatic image interpretation associated to the management of many diseases. Precise measurement of the morphological changes of these curvilinear organ structures informs clinicians for understanding the mechanism, diagnosis, and treatment of e.g. cardiovascular, kidney, eye, lung, and neurological conditions.  In this work, we propose a generic and unified convolution neural network for the segmentation of curvilinear structures and illustrate in several 2D/3D medical imaging modalities. We introduce a new curvilinear structure segmentation network (CS$^2$-Net), which includes a self-attention mechanism in the encoder and decoder to learn rich hierarchical representations of curvilinear structures. Two types of attention modules  - spatial attention and channel attention - are utilized to enhance the inter-class discrimination and intra-class responsiveness, to further integrate local features with their global dependencies and normalization, adaptively. Furthermore, to facilitate the segmentation of curvilinear structures in medical images, we employ a 1$\times$3 and a 3$\times$1 convolutional kernel to capture boundary features. Besides, we extend the 2D attention mechanism to 3D to enhance the network’s ability to aggregate depth information across different layers/slices. The proposed curvilinear structure segmentation network is thoroughly validated using both 2D and 3D images across six different imaging modalities. Experimental results across nine datasets show the proposed method generally outperforms other state-of-the-art algorithms in various metrics.
\end{abstract}


\end{frontmatter}


\section{Introduction}

\textit{Curvilinear structures} are objects with thin, long, elongated and, sometimes, arborescent shape, and present distinct intensity when compared to their neighbouring structures ~\citep{Bibiloni}. In the biomedical field, many studies~\citep{Kim2018,rieber2006cardiac} suggest geometrical and topological changes in numerous anatomical curvilinear structures - e.g., retinal blood vessels, cerebral vasculature, lung airways, or nerve fibres - are closely linked to the presence or severity of  diseases, including, for instance, diabetes, stroke, hypertension, and keratitis.

\begin{figure*}[t]
	\centering
	\includegraphics[width=0.98\textwidth]{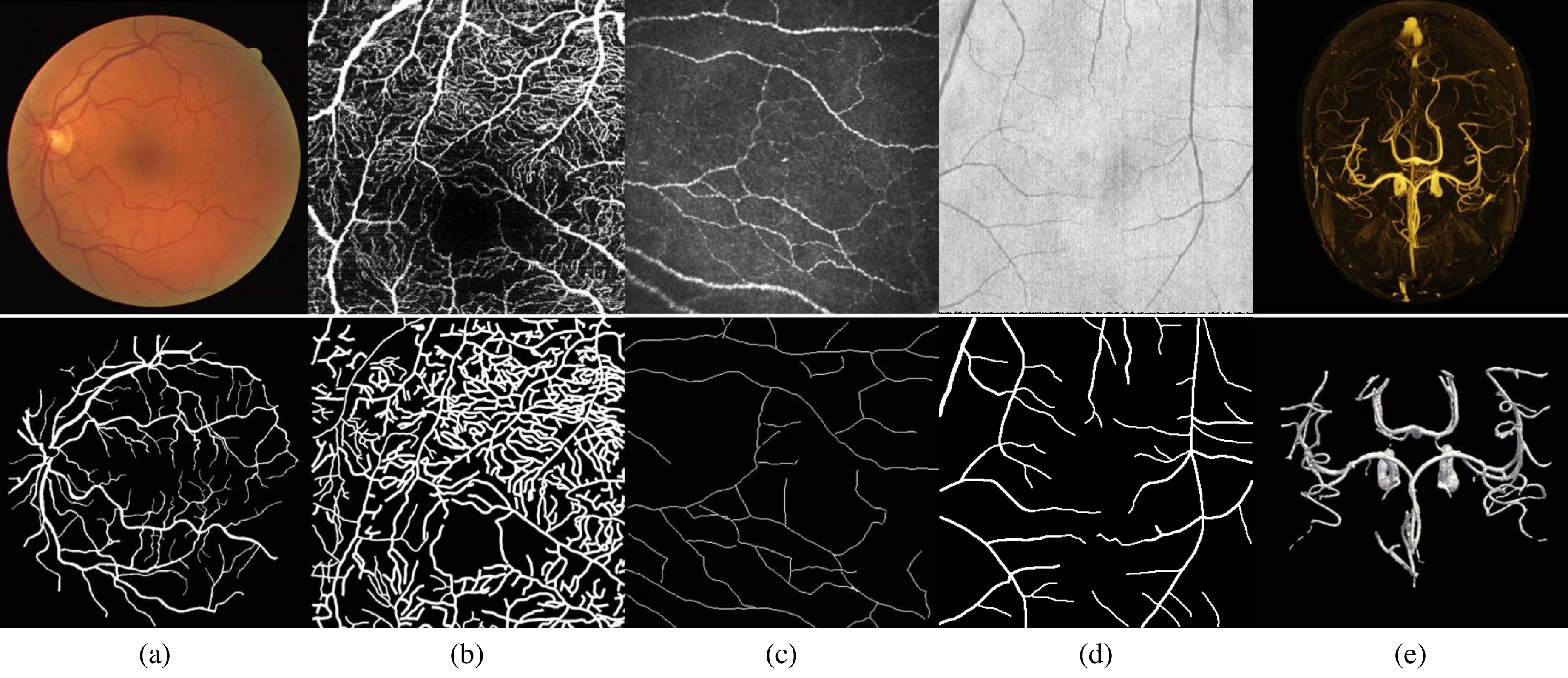}
	\caption{Images (tow row) and their manual annotations of curvilinear structures (bottom row) in different medical imaging modalities. From the left to right column: Retinal color fundus image; Retinal optical coherence tomography angiogram (OCTA); Corneal confocal microscopy (CCM) image; Optical coherence tomography (OCT) and Brain MRA. Note that the manual annotations of OCTA and CCM are made at a centerline level, and the cerebral vasculatures  are visualized in 3D by maximum intensity projection. 
	}
	\label{fig1}
\end{figure*}

Acquiring images of these anatomical curvilinear structures has impact in a number of two-dimensional (2D) and three-dimensional (3D) imaging modalities, such as colour fundus imaging, optical coherence tomography angiography (OCTA), fluorescence angiogram (FA), confocal microscopy (CM), magnetic resonance angiography (MRA), computed tomography angiography (CTA), etc. The top row of Fig.~\ref{fig1} demonstrates five examples of different medical image types, which include both 2D (Fig.~\ref{fig1}(a-d)) and 3D (Fig.~\ref{fig1}(e)) images.

As one type of curvilinear structure, retinal blood vessels are an essential component of the retina, and the morphology change of  their  retinal  vasculature  is closely related to many systemic, metabolic, and haematologic diseases~\citep{annunziata2016fully, ding2014supervised}. Retinal blood vessels are usually observed in colour fundus images~\citep{franklin2014computerized} and OCTA images~\citep{Carlo}. Colour fundus imaging can exclusively reveal the superficial vascular network. At the same time, OCTA is a new, non-invasive imaging technique that generates volumetric angiography images, and can visualise the radial peripapillary and deep capillary networks that are not well-distinguished in colour fundus images. Corneal nerve fibre properties such as branching, density, and tortuosity are linked to eye and systemic diseases such as herpes, simplex keratitis and dry eye diseases~\citep{eladawi2017automatic, Kim2018}. In vivo corneal confocal microscopy (CCM) is a common technique for the imaging and inspection of corneal nerve fibres. Early detection of their geometrical and topological changes often helps to reduce the incidence of vision loss and blindness. MRA is an MRI examination of the human brain vessels (cerebral vasculature), which is vital for the diagnosis of many serious diseases such as strokes~\citep{liao2012automatic}. Cerebral small vessel deformation plays an indicative role in lacunar strokes and brain haemorrhages and are a leading cause of cognitive decline and functional loss in elderly patients~\citep{cuadrado2018cerebral}. 

In consequence, accurate extraction of these curvilinear structures from medical images is often an essential step in quantitative image analysis and computer-aided diagnostic pipelines. The bottom row of Fig.~\ref{fig1} illustrates the manual annotations of five types of medical images. However, manual annotation of these curvilinear structures is an exhaustive time-consuming task for graders, and subject to human error,  and thus impractical in high-throughput analysis settings like screening programmes or  microscopy~\citep{ZhaoTMI18}.  In addition, the commercial software available (e.g. ImageJ{\footnote{\url{https://imagej.nih.gov/ij/}}}and TubeTK{\footnote{\url{http://tubetk.org/}}}) still rely heavily on manual refinement. This calls for fast, accurate, and fully automated curvilinear structure extraction methods. 

Over the last two decades, we have witnessed the rapid development of curvilinear structure detection methods, especially for blood vessel segmentation, as evidenced by general reviews of 2D vessel segmentation~\citep{Fraz,ZhaoTMI18}, and 3D vessel segmentation~\citep{Lesage09}. Most existing segmentation methods suffer from issues posed by high anatomical variability across populations, and the varying scales of curvilinear structures within an image. On one hand, noise, poor contrast and low resolution exacerbate these problems. Standard image segmentation methods often cannot robustly detect all the curvilinear structures of interest.  On the other hand, deep learning-based techniques have yet to be used to segment retinal vessels in OCTA and most of them are designed for the segmentation of vessels or fibers from one specific biomedical imaging modality. Moreover, most of them are designed specifically for 2D images and cannot easily be extended to 3D ones. It has proven very challenging to develop a single curvilinear structure detection method that works well across a variety of medical imaging modalities.

In this paper, we introduce a novel Channel and Spatial Attention Network (CS$^2$-Net) to extract curvilinear structures from images in different imaging modalities. Our work was inspired by Dual Attention Network (DANet)~\citep{fu2018dual} that were designed for the segmentation of natural images. While medical images contain more unique features, such as simpler semantics and unitary patterns, we first construct a network backbone based on the encoder-decoder framework, and then introduce a $1\times 3$ and a $3 \times 1$ convolutional kernel to capture more boundary feature to assist the segmentation of curvilinear structures, rather than only up-sampling the attention features in the last layer of DANet. Such approach is more attractive to the researchers and practitioners, since they need not choose a particular method for each imaging modality and it is more applicable to various imaging modalities. The proposed method extends considerably our previous work~\citep{MouMICCAI}, which focused on 2D curvilinear structure segmentation in medical images only. In this work, we have improved it so that it is applicable to segment the curvilinear structures from both 2D and 3D imaging modalities. We have also expanded our data pool for evaluation from three biomedical imaging modalities to six with a total of nine different datasets. Overall, this work makes the following contributions:
\begin{enumerate}
	\item[1)] A new curvilinear structure segmentation network is proposed based on dual self-attention modules, which can deal with both 2D and 3D imaging modalities in an unified manner; 
	\item[2)] Two self-attention mechanisms are employed in the channel and spatial spaces to generate attention-aware expressive features. They can enhance the network to capture long-range dependencies and make an effective use of the multi-channel space for feature representation and normalization, enabling the network to classify the curvilinear structure from background more effectively;
	\item[3)] Experimental results on nine datasets (six 2D datasets and three 3D datasets) demonstrate that our proposed CS$^2$-Net achieves on the whole state-of-the-art performances in detecting curvilinear structures from different biomedical imaging modalities both quantitatively and qualitatively. The source code of this work is available at: \url{https://github.com/iMED-Lab/CS-Net}
\end{enumerate}

\section{Related Works}
\subsection{2D Curvilinear Segmentation}


As vessels, airways or fibres in 2D medical images are curvilinear structures distributed across different orientations and scales, various filtering methods have been proposed, including Hessian matrix-based filters~\citep{Frangi}, 
matched filters~\citep{Zhao2017Neuro}, 
multi-oriented filters~\citep{soares2006retinal,zhang2017retinal}, symmetry filter~\citep{ZhaoTMI18}, and tensor-based filter~\citep{Cetin2015}, 
active contours-based methods~\citep{Shang2011Vascular, Al-Diri2009Active} and minimal geodesic paths-based approaches~\citep{ChenTIPMinimal}. These filtering-based methods aim to suppress non-vascular or non-fiber structures and imaging noise, and enhance the curvilinear structures, thereby benefiting the subsequent segmentation problem. For instances,~\citep{zhao2015automated, zhao2018automatic} proposed infinite perimeter active contour model with hybrid region information and a weighted symmetry filter to detect vessels. \citep{zhang2016robust} designed multi-scale rotation invariant filters for retinal vessel and corneal nerve fibre segmentation based on a locally adaptive framework in the position and orientation spaces. This framework is adaptive to the local changes of curvilinear structures and can deal with typically difficult cases. \citep{soares2006retinal} used a multi-scale Gabor transform to extract texture features of vessels for more accurate vessel detection. There are also several filter-based vessel segmentation methods, including Hessian matrix-based filters~\citep{Frangi,zhang2016robust}, tensor-based filters~\citep{Cetin2015} and symmetry filters~\citep{ZhaoTMI18}. These approaches aim to remove undesired intensity variations in the images, and suppress background structures and imaging noise, thereby facilitating the subsequent segmentation task. However, these filter-based methods usually rely heavily on manual parameter adjustment during implementation, and are designed mainly for a specific imaging modality, which may not be effective when applied to other image types.

Recently, deep learning-based methods have made significant progress in computer vision. These include classification networks, e.g., ResNet~\citep{he2016deep} and Inception series networks~\citep{szegedy2015going, ioffe2015batch,szegedy2016rethinking,szegedy2017inception};  object detection networks, e.g., Faster-RCNN~\citep{ren2015faster} and R-FCN~\citep{dai2016r};  segmentation networks, e.g., SegNet~\citep{badrinarayanan2017segnet}, PSPNet~\citep{zhao2017pyramid}; and networks designed for medical image segmentation, e.g., U-Net~\citep{ronneberger2015u} and CE-Net~\citep{Gu2019CENetCE}. These deep learning based methods have been modified and applied for blood vessel segmentation~\citep{Fu2016DeepVesselRV,alom2018recurrent} 
and nerve fibre tracing in colour fundus and CCM images~\citep{colonna2018segmentation,williams2020} , respectively. \citep{Maninis16} proposed a multi-task structure for both vessel detection and optic disc segmentation. \citep{Liskowski2016} introduced a retinal vessel segmentation method based on a convolutional neural network (CNN), and \citep{Fu2016DeepVesselRV} further applied the CNN along with conditional random fields for the detection of retinal vessels. \citep{alom2018recurrent} embedded a recurrent neural network into the U-shaped network (R2U-Net) for the segmentation of vessels. \citep{wang2019high} proposed a novel detector, named Oriented Cylinder Flux (OCF), for the detection of blood vessel structures. \citep{wang2019context} proposed a new curvilinear structure segmentation method using context-aware and spatio-recurrent networks. Instead of directly segmenting the entire image or densely segmenting fixed-size local patches, it uses a learning strategy to sample the target image with different proportions repeatedly. More details on recent vessel segmentation works can be found in~\citep{shin2019deep, jin2019dunet, wang2019blood}. \citep{colonna2018segmentation} proposed a deep neural network based on U-Net~\citep{ronneberger2015u} for corneal fibre tracing in CCM images. \citep{hosseinaee2019fully} developed an automated method for the segmentation of corneal nerves on en face UHR-OCT images obtained from healthy human subjects. \citep{kim2018automatic} and \citep{oakley2019deep} systematically summarized the unsupervised and supervised methods for corneal nerve segmentation and analysed the role of corneal neuromorphological features in disease diagnosis. \citep{eladawi2017automatic}~proposed a joint Markov-Gibbs random field (MGRF) model to segment blood vessels based on different retinal maps from OCTA scans. \citep{diaz2019automatic} developed an automatic system that identifies and precisely segments the foveal avascular zone (FAZ). \citep{heisler2019deep} also proposed a novel automated deep learning method to segment and quantify retinal images from prototype OCTA machines with larger fields of view. For more automated vascular segmentation and fibre tracing methods, please refer to the review by~\citep{Fraz}.
Although these methods have achieved promising segmentation results,  most of them concentrate on the segmentation of curvilinear structure for single imaging modality. Besides, most of them are hard to be extended for the curvilinear structure segmentation in 3D volumes.

\subsection{3D Curvilinear Segmentation}

Three-dimensional volumes contain richer features with depth information not available in 2D slices/images. Three-dimensional vascular segmentation is an essential prior step in the characterization of cerebral aneurysms, which has proven useful for the pre-treatment planning of Guglielmi separable coils (GDC)~\citep{wilson1998improved}. 
With developing imaging devices, more computer vision methods have been developed to deal with 3D data for biomedical data analysis. Most methods perform better on this volumetric data compared to 2D image counterpart, especially in medical imaging. \citep{ZhaoTMI18} proposed a weighted symmetry filter for automatic 2D vessel enhancement and segmentation, and further extended it to the 3D case for vascular segmentation. 
\citep{unet3d} extended U-Net to 3D U-Net with a weighted cross entropy loss to perform Xenopus kidney segmentation, which has been proven an effective method for the segmentation of tubular structural organs under sparse annotations. \citep{gibson2018automatic} used a dense V-Net~\citep{milletari2016v} to segment multiple 3D tubular organs. \citep{chung1999statistical} adopted a Rician distribution to segment 3D brain vasculatures in order to extract cerebral aneurysm features. \citep{DeepVesselNet} proposed DeepVesselNet to segment vessels, detect vessel centerlines and bifurcate 3D angiographic volumes. \citep{liao2012automatic} recommended to segment human brain vessels  using fast matching with an anisotropic orientation being a priori. Recently, \citep{zhang20193d} proposed a novel method for 3D retinal OCTA microvascular segmentation and surface reconstruction. Intrinsic shape analysis was performed to extract useful surface-based 3D geometric and topological biomarkers. \citep{wang2019segmenting} proposed a teacher-student learning framework for fast neuron segmentation, where the segmentation inference is performed using a light-weighted student network which benefits from knowledge distillation by a teacher network with a higher capacity. \citep{zhao2019deep} proposed to perform 3D vessel segmentation by utilizing a deep feature regression (DFR) method based on a convolutional regression network (CRN) and a stable point clustering mechanism. \citep{poulain20193d} proposed a new approach by combining the information of a tree-spline with a registration algorithm to perform 3D coronary vessel tree tracking. \citep{uception} proposed a Uception network based on Inception modules and the U-Net-like architecture to segment cerebrovascular in MRA images. However, many modules that rely heavily on GPU resources are used in Uception, which makes the method require considerable GPU memory resources during training and inference stages.

Like the previous 2D segmentation methods, many filter-based 3D tubular structure segmentation methods rely heavily on manual tuning. Some methods based on learning strategies ignore the tubular structure by designing particular network modules, which plays a vital role in their accurate segmentation.

\section{Proposed Method}
\subsection{Network Architecture}
The proposed CS$^2$-Net is designed for curvilinear structure segmentation of both 2D and 3D medical images. It consists of three modules: the encoder module, the channel and spatial attention module (CSAM), and the decoder module. Fig.~\ref{fig-csnet2d} and Fig.~\ref{fig-csnet3d} illustrate the architectures for 2D and 3D images, respectively.
The  encoder module is used to extract the features of input data. Then, these features are fed into two parallel attention blocks - the channel attention block (CAB) and a spatial attention block (SAB) - to generate channel-spatial attention-aware expressive features. The SAB selectively aggregates the features in each spatial location through the weighted features in all spatial locations, which allows the model to capture the long-range dependency of the features, and similar features will be related to each other regardless of their distance. Meanwhile, the CAB makes sure that the full space is used to represent and normalize and thus enhance the contrast of the features in different channels, allowing the model to be assembled with improved discrimination capabilities. Finally, the decoder module is employed to reconstruct curvilinear features and produce the segmentation result.

Instead of directly up-sampling the features of the CSAM to the original image dimensions~\citep{fu2018dual}, we introduce a feature decoder module that restores the dimensions of the high-level semantic features layer by layer. The encoder and decoder modules include four blocks, each of which employs a residual network (ResNet) as the backbone, and then followed by a max-pooling layer to increase the receptive field for better extraction of global features. Similar to the U-shaped network~\citep{ronneberger2015u,unet3d}, a skip connection between each layer of the encoder and decoder is introduced to combine the features at different levels to compensate for information loss caused by the max-pooling operations. At the end of the CS$^2$-Net, we apply a $1 \times 1$ kernel ($1 \times 1 \times 1$ kernel in the 3D phase) convolutional layer and a sigmoid layer on the output of the encoder to obtain the final segmentation map. 

\subsection{2D Attention Network}
\begin{figure*}[!t]
	\centering
	\includegraphics[width=0.9\textwidth]{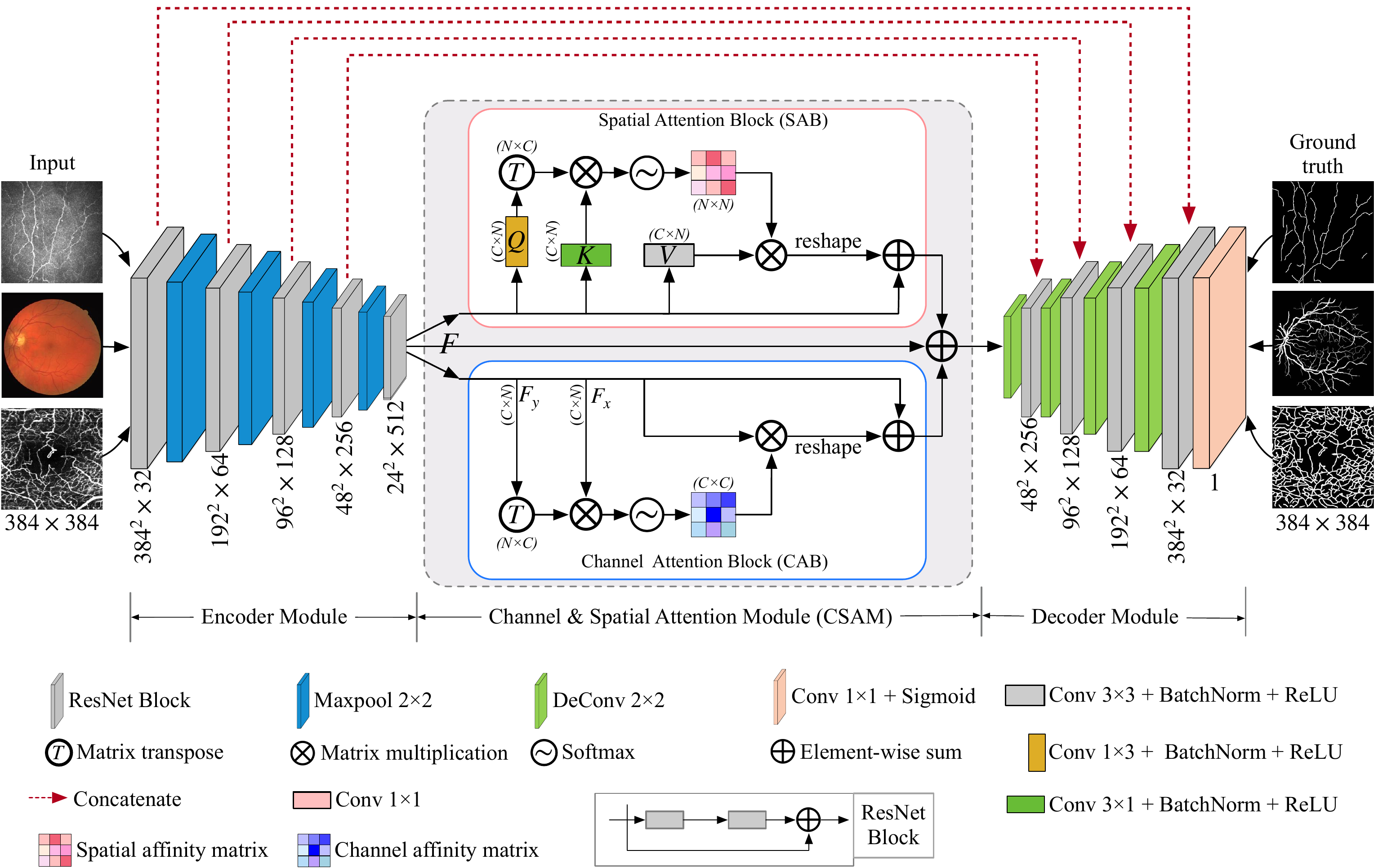}
	\caption{The architecture of the proposed CS$^2$-Net over the 2D images: an encoder, a CSAM module and a decoder, which extract global features, enhance the feature expression ability, and reconstruct curvilinear features, respectively.}
	\label{fig-csnet2d}
\end{figure*}

Several recent works have shown that the local feature representations produced by traditional fully convolutional networks (FCNs) may lead to object misclassification~\citep{zhao2017pyramid,peng2017large}. The CS$^2$-Net, which consists of a 2D encoder, 2D CASM and 2D decoder, reduces this limitation and segment curvilinear structures in 2D images more effectively. The 2D version CSAM is shown in Fig.~\ref{fig-csnet2d}, which includes a 2D SAB and a 2D CAB. We use 2D convolutional, 2D batch normalization and 2D deconvolutional layers in all the modules. Their working principles are explained as follows.

\subsubsection{2D Spatial Attention Module}
To model rich contextual dependencies over local feature representations, the first step is to generate a spatial attention matrix, which models spatial relationships between the features of any two pixels. Tree-like structures are always distributed throughout the biomedical images. Following~\citep{fu2018dual}, we modify the SAB to encode a broader range of contextual information about local features, and increase their representation capability. However, unlike~\citep{fu2018dual}, we introduce a $3 \times 1$ and a $ 1 \times 3$ convolutional layer with batch normalization and ReLU layers to capture the edge information of the tree-like structures in horizontal and vertical orientations, respectively. More importantly, compared with many complex natural images, medical images contain rare and almost fixed structures. Considering this aspect, the curvilinear structure segmentation network requires skip-connection operations to fuse low-level information and compensate for the lost spatial information. Therefore, we transplant the proposed attention module into the encoder-decoder framework, rather than directly encoding the image and resampling the original one as in~\citep{fu2018dual}.

Specifically, we place the two types of layers ($3 \times 1$ and $1 \times 3$ convolutional layer) after the input features $F \in \mathbb{R}^{C\times H\times W}$ to generate two new feature maps $Q_y \in \mathbb{R}^{C\times H\times W}$, and $K_x \in \mathbb{R}^{C\times H\times W}$, respectively, where $C$ denotes the dimensionality of the input features, $H$ and $W$ are the height and width of the input image, $Q_y$ and $K_x$ represent the features of the curvilinear structures captured in the vertical and horizontal directions. These two new feature maps are then reshaped to $\mathbb{R}^{C\times N}$, where $N=H\times W$ is the number of features. In consequence, the intra-class spatial association can be obtained by applying a softmax layer on the matrix multiplication of the transpose of $Q$ and $K$, as:
\begin{equation}
\mathcal{S}_{(x,y)}=\frac{\exp{\left( {Q^{\rm T}}_{y} \cdot {K}_{x} \right)}}{\sum_{x'=1}^N \exp{\left( {Q^{\rm T}}_{y} \cdot {K}_{x'} \right)}} ,
\end{equation}
where $\mathcal{S}_{(x,y)}$ denotes the $y^{th}$ position's impact on the $x^{th}$ position. Matrix multiplication computes and outputs the feature correlation matrix $\mathcal{S}_{(x,y)}$ between any two points, the two similar spatial points promote each other and the two different spatial points suppress each other. Through this operation, the network can fully utilize and learn the curvilinear structure of different spatial locations. Then we apply softmax on the correlation matrix to obtain the attention map of the similarity between each spatial position and the others, in which the higher the similarity, the greater the response between the two points. Meanwhile, another new feature $V \in \mathbb{R}^{C\times H\times W}$ is obtained by applying a $1 \times 1$ convolutional layer with batch normalization and ReLU layers on the input features and we also reshape it to $\mathbb{R}^{C \times N}$, which is then used to perform a matrix multiplication with $\mathcal{S}_{(x,y)}$ to obtain the attention enhanced features $F' \in \mathbb{R}^{C\times N}$. Finally, we reshape it to $\mathbb{R}^{C \times H \times W}$, and perform channel-wise addition of $F$ and $F'$ over each pixel to construct the output of SAB. Thus, SAB gains a global contextual view and selectively aggregates context information according to the spatial attention map to achieve a more accurate segmentation performance for curvilinear structures.

\subsubsection{2D Channel Attention Module}
Since each channel of a high-level feature can be regarded as a specific-class response, we further exploit the inter-dependencies of channel maps in this section, and propose the CAB module to improve the feature representation by using the space available. 
A channel-wise attention map is obtained by applying a softmax layer on the channel-wise similarity map between the input feature $F$ (named as $F_x \in \mathbb{R}^{C\times H \times W}$) and its transpose (named as ${F^{\rm {T}} \in \mathbb{R}^{H \times W \times C}}$) as:
\begin{equation}
\label{eq1}
\mathcal{C}_{(x,y)}=\frac{\exp{\left( {F}_{x} \cdot {F}_{y}^{\rm T} \right)}}{\sum_{x'=1}^{C}\exp{\left( {F}_{x'} \cdot {F}_{y}^{\rm T} \right)}},
\end{equation}
where $\mathcal{C}_{(x,y)}$ denotes the attention of the $x^{th}$ channel relative to the $y^{th}$ channel. Therefore, we can obtain the channel dependency matrix ($\mathbb{R}^{C \times C}$, where $C$ denotes the number of channels) by performing a matrix multiplication. Here, two similar channels will promote each other. In contrast, different channels will inhibit each other. After that, a softmax is applied on the channel dependency matrix to enhance the discrimination between curvilinear structure and its background. The process is similar to the spatial attention module above. The difference lies in two aspects: (i) while the former operates the original features $F$ directly, the latter works on newly derived features $Q_y$, $K_x$ and $V$, and (ii) while the former models the attention of the features in one channel relative to those in another, the latter models the attention of features at one pixel relative to those at another. Similar to SAB, we then perform a multiplication between $\mathcal{C}_{(x,y)}$ and $F$ to obtain the attention enhanced features $F''$. The final output of CAB is defined as $F+F''$ over each pixel. Such operations enhance the contrast between class-dependent features and help improve their expressiveness.

\begin{figure*}
	\centering
	\includegraphics[width=0.9\textwidth]{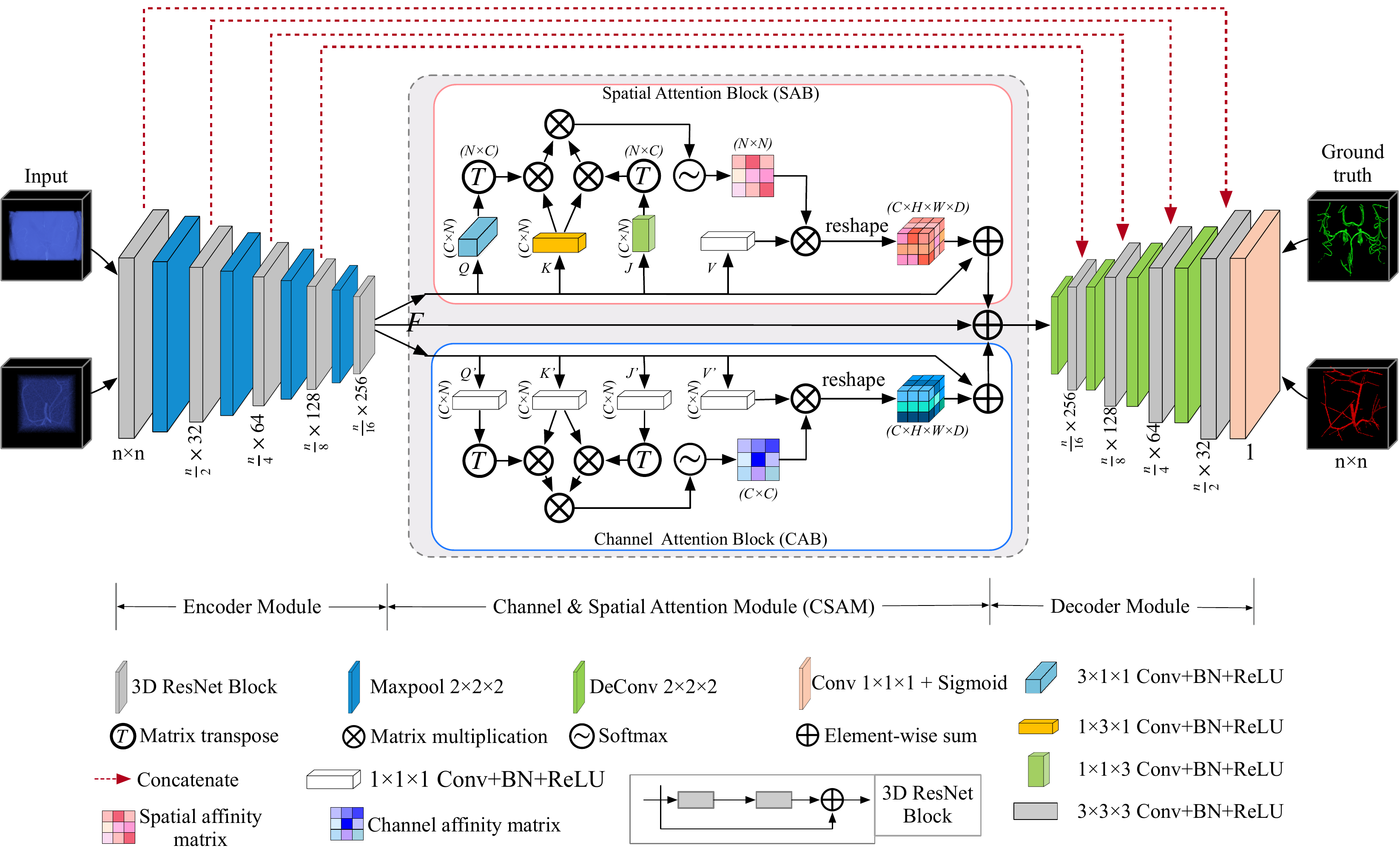}
	\caption{The diagram of the 3D CS$^2$-Net.  It includes an  encoder,  a 3D CSAM and a decoder.  N is the size of cropped volumes during training.}
	\label{fig-csnet3d}
\end{figure*}

\subsubsection{Objective Function}
All datasets contain complete  annotations, and curvilinear structure segmentation in a 2D image can be regarded as a pixel-level binary classification task: curvilinear structure or background. In this work, the binary cross-entropy (BCE) loss is thus adopted as the objective function for the training of the network, as it is a pixel-wise objective function that directly evaluates the distance between the ground truth and prediction. The BCE loss is defined as:
\begin{equation}
\nonumber
\mathcal{L}_{BCE}=-\frac{1}{N}\sum_{i=1}^N g_i\cdot \log(p_i)+(1-g_i)\cdot \log(1-p_i),
\end{equation}
where $g_i \in \{0,1\}$ indicates the ground truth as curvilinear structure of a pixel, $p_i \in [0,1]$ is its predicted probability, and $N$ is the number of pixels.

\subsection{3D Attention Network}
In recent years, many methods based on learning and manual design have been proposed for the detection of curvilinear structures in 2D images~\citep{staal2004ridge, Kim2018, li2015automated}. However, there are relatively few methods, especially learning-based methods, for segmenting curvilinear structures in 3D images. Moreover, the spatial attention and the channel attention in~\citep{fu2018dual} focus only on the 2D domain. Directly applying a 2D attention on  3D images lacks feature integration in the depth direction, which is crucial for improving the results of segmentation of curvilinear structures. To enable our proposed CS$^2$-Net to extract the 3D tree-like structures, we extend it from the 2D to the 3D, as shown in Fig.~\ref{fig-csnet3d}. For the encoder and decoder modules, we replace all their 2D operations with 3D ones. However, due to changes in the modality of the dataset, the proposed CSAM in the 3D mode differs significantly from that in the 2D one. We detail the 3D CSAM in the following section.

\subsubsection{3D Spatial Attention Module}
Similar to the 2D SAB, we first feed the input features $F \in \mathbb{R}^{C\times H \times W \times D}$ into a $1\times 3 \times 1$ and $3\times 1 \times 1 $ layer with batch normalization and ReLU activations to generate two feature maps $Q_y\in \mathbb{R}^{C\times H \times W \times D}$ and $K_x\in \mathbb{R}^{C\times H \times W \times D}$ to capture the boundary features of tublar structure along $y$ axis and $x$ axis, where $C$ indicates the number of input channels, and $H,W$ and $D$ indicate the height, width and depth of the input 3D image, respectively. 
However, this operation encodes the relationship solely between features in the width and height directions, lacking feature integration in the depth direction. To overcome this limitation, we also feed $F$ into a $1 \times 1 \times 3$ convolutional layer and then optimize and activate it with batch normalization and ReLU layers. A new feature map $J_z\in \mathbb{R}^{C\times H \times W \times D}$ is obtained. Therefore, we use $Q_y$, $K_x$ and $J_z$ to capture the edge information of tree-like structures in the width, height and depth directions. In the next step, we reshape these three feature maps to $\mathbb{R}^{C\times N}$ to construct three activation matrices, where $N= H \times W \times D$. We first perform matrix multiplication on $Q_y^{\rm T}$ and $K_x$ to encode the feature relationships in the width and height directions, and then operate $K_x$ and $J_z^{\rm T}$ to encode the feature relationships in the height and depth directions, where $Q_y^{\rm T}$ and $J_z^{\rm T}$ are the transpose of $Q_y$ and $J_z$, respectively. To encode the relevance of features in the width and depth directions, we apply a matrix multiplication between the two outputs from the previous step. Finally, a softmax layer is used to obtain the voxel-level, intra-class affinities as:
\begin{equation}
\mathcal{S}_{(x,y,z)}=\frac{\exp \left[ \left( Q_{y}^{\rm T} \cdot K_{x} \right) \cdot \left( J_{z}^{\rm T}\cdot K_{x} \right)\right] }{\sum_{x'=1}^N \exp \left[ \left( Q_{y}^{\rm T} \cdot K_{x'} \right) \cdot \left( J_{z}^{\rm T} \cdot K_{x'} \right)\right] },
\end{equation}
where $\mathcal{S}_{(x,y,z)}$ denotes the mutual impacts of features at the $x^{th}$, $ y^{th}$ and $z^{th}$ positions. Similarly, we gain a dimension-reduced feature map $V \in \mathbb{R}^{C\times H \times W \times D}$ by applying a $1 \times 1 \times 1$ kernel convolutional layer on the input feature map $F$, and we reshape it to $\mathbb{R}^{C \times N}$. Then, a matrix multiplication is performed between $V$ and $\mathcal{S}_{(x,y,z)}$ to obtain the voxel-level attention enhanced features $F'$, which is then reshaped to $\mathbb{R}^{C \times H \times W \times D}$. 
Finally, we add $F'+F$ channel-wise over each voxel to obtain the output of the 3D SAB. The schematic diagram of CSAM in Fig.~\ref{fig-csnet3d} shows the details of this process. Our proposed 3D SAB not only performs feature mapping in the width and height directions, but also performs the mutual mapping of the 3D features in the depth direction. It is expected to increase the feature expression ability of the network.

\subsubsection{3D Channel Attention Module}
Inspired by the 2D channel attention mechanism, we further extend it to the 3D domain. Similar to the 2D CAB, we apply a $1\times 1 \times 1$ kernel convolutional layer on $F \in \mathbb{R}^{C\times H \times W \times D}$ to derive four new feature maps ${Q'}_y \in \mathbb{R}^{C \times H \times W \times D}$, ${K'}_x \in \mathbb{R}^{C \times H \times W \times D}$, ${J'}_z \in \mathbb{R}^{C \times H \times W \times D}$ and $V' \in \mathbb{R}^{C\times H \times W \times D}$, respectively. Then, we reshape ${Q'}_y$, ${K'}_x$, ${J'}_z$ and $V'$ to $\mathbb{R}^{C \times N}$. Finally we perform the same matrix operations on ${Q'}_y$, ${K'}_x$ and ${J'}_z$ as in the 3D SAB:
\begin{equation}
\mathcal{C}_{(x,y,z)}=\frac{\exp \left[ \left( {K'}_{x} \cdot {Q'}_{y}^{\rm T} \right) \cdot \left( {K'}_{x} \cdot {J'}_{z}^{\rm T} \right) \right]}{\sum_{x'=1}^C \exp \left[ \left( {K'}_{x'} \cdot {Q'}_{y}^{\rm T} \right) \cdot \left( {K'}_{x'} \cdot {J'}_{z}^{\rm T} \right) \right]},
\end{equation}
where $\mathcal{C}_{(x,y,z)}$ denotes the mutual affinities between the $x^{th}$, $y^{th}$ and $z^{th}$ channels. Besides, a matrix multiplication is performed between the transpose of $\mathcal{C}$ and $V'$  to obtain the voxel-level channel-wise attention enhanced features $F''$ and reshape it to $\mathbb{R}^{C \times H \times W \times D}$. Similarly, we add $F''$ and $F$ channel-wise over each voxel to obtain the output of the 3D CAB.

To gather the spatial and channel attention maps, a voxel-level matrix summation is applied as the output of the 3D CSAM between the outputs of the 3D SAB and the 3D CAB and the original input feature $F$. 

\subsubsection{Loss Function}
The labels for 3D cerebrovascular regions are sparse, and only a portion of them have high-quality annotations. Thus, we choose as our loss function the weighted cross entropy loss $L_{WCE}$ (WCE), which can adjust learning bias between a vascularity and background during training. Moreover, we also introduce Dice coefficient loss $L_{Dice}$ to ensure the micro-cerebrovascular segmentation. Finally, we define the 3D optimization loss function for the training of the proposed $CS^2$-Net as:
\begin{equation}
\mathcal{L}=\alpha L_{WCE} +(1-\alpha) L_{Dice},
\end{equation}
where $\alpha$ is the weight balance parameter between $L_{MSE}$ and $L_{Dice}$, which is empirically set as $\alpha=0.6$. For our binary segmentation task, the WCE loss  and the Dice coefficient loss are defined:
\begin{equation}
L_{WCE}=-\frac{1}{N}\sum_{i=1}^{N} \left(\omega g_i \log p_i + \left( 1-g_i \right) \log \left( 1-p_i \right)\right), 
\end{equation}
\begin{equation}
L_{Dice}=1-\frac{2\sum_{i=1}^{N}p_i g_i+\epsilon}{\sum_{i=1}^{N}p_i^2+\sum_{i=1}^{K}g_i^2+\epsilon},
\label{mseloss}
\end{equation}
where $\omega$ is the class weight of curvilinear structure, and can be obtained by the class estimation probabilities $p_i$ of all the voxels: 
$$ {\omega}=\frac{N-\sum_{i=1}^N p_i}{\sum_{i=1}^N p_i}.$$
Here, $N$ denotes the number of voxels, and $p_i \in [0,1]$ and $g_i \in \{0,1\}$ denote the predicted probability and ground truth value of the $i^{th}$ voxel as the curvilinear structure, respectively. The parameter $\epsilon$ is a Laplace smoothing factor used to avoid numerical instability problem and accelerate the convergence of the training process ($\epsilon=1.0$ in this paper).



\begin{table}[h]
	\centering
	\caption{Details of the datasets  used to evaluate the proposed method.}
	\label{tab-2ddata}
	\begin{tabular}{lccll}
		\hline
		\hline
		Datasets & Number & Resolution & Data type & Public \\
		\hline
		DRIVE & 40 & $565 \times 584$ &  Fundus & Public  \\
		STARE & 20 & $605 \times 700$ &  Fundus & Public   \\
		IOSTAR & 30 & $1024 \times 1024$ &
		Fundus & Public  \\
		CORN-1 &  1698 & $384 \times 384$ &  CCM & Public  \\
		OCTA & 30 &  $1376 \times 968$ &  OCTA & Private  \\
		OCT RPE  & 36 & $384 \times 379$ &  OCT & Private \\
		\hline
	\end{tabular}
\end{table}

\section{Experimental Results over 2D Images}

In this section, the proposed segmentation network is first validated over 2D medical images for the extraction of their curvilinear structures. Many datasets are available online and aim to train and validate an automatic approach for the segmentation of vessels or nerve fibres from 2D medical images, as blood vessels or nerve fibres are closely correlated to the presence of pathology. We refer readers to~\citep{ZhaoTMI18} for more detailed introduction and discussions. In this work, we selected two most commonly used (DRIVE and STARE), two newly released (IOSTAR and CORN-1) publicly available datasets, and two private (OCTA and OCT RPE) datasets for evaluation of our method and the competitors.

In this work, we selected the most commonly used datasets in the research community to evaluate the proposed CS$^2$-Net, so that we can make a direct comparison of segmentation results with those obtained by the state-of-the-art methods. Regarding the two private datasets, OCT and OCTA are two new emerging non-invasive imaging techniques, with the ability to produce high-resolution 3D images of retinal vasculature, and have been increasingly taken as a valuable imaging tool to observe retinal vascular. To our best knowledge, there is no publicly available OCTA or OCT 
RPE dataset with manually graded vessels for training and validation. We use these two datasets to test our model, keep growing in the size of these two datasets, and will release them online in the future.

\subsection{Materials}
Six 2D datasets in total are used for evaluation, whose details are provided   as follows.

\textbf{DRIVE}{\footnote{\url{http://www.isi.uu.nl/Research/Databases/DRIVE/}}} contains 40 colored fundus images, which were initially divided into 20 images for training and 20 images for testing. The images were acquired using a Canon CR5 non-mydriatic 3-CCD camera with a field of view (FOV) being  $45^\circ$. Each image in this dataset has dimensions of $565 \times 584$. We follow the same partition of the images in our training and testing.

\textbf{STARE}{\footnote{\url{http://www.ces.clemson.edu/ ahoover/stare/}}} comprises 20 colored fundus images. The images were captured using a Topcon TRV-50 fundus camera with a FOV being $35^\circ$. Half of the images contain pathological indications and the other half come from healthy subjects. Each image has dimensions of $700 \times 605$. However, unlike the DRIVE dataset above, there is no fixed partition of training and testing sets. In this paper, we adopt the k-fold (k=4) cross-validation method for the training and testing phases, similar to that in~\citep{mo2017multi}. Therefore, 15 images are used for training and the remaining 5 images are used for testing in each fold. We use the manual annotations from the first observer as the ground truth for all the images.

\textbf{IOSTAR}{\footnote{\url{http://www.retinacheck.org/}}} includes 30 images with a resolution of $1024 \times 1024$ pixels. The images were acquired with an EasyScan camera (i-Optics Inc., the Netherlands), which is based on a SLO technique with a FOV being 45 degrees. For reasonable data division, we also adopt the k-fold (k=5) cross-validation method for training and evaluating, that is, 24 images are used for training and 6 images for testing.

\textbf{CORN-1}{{\footnote{\url{http://imed.nimte.ac.cn/}}}} is a publicly available CCM dataset, and contains a total of 1698  CCM images of corneal subbasal epithelium using a Heidelberg Retina Tomograph equipped with a Rostock Cornea Module (HRT-III) microscope. These images were  acquired by the Peking University Third Hospital, China and  University of Padova, Italy{{\footnote{\url{http://bioimlab.dei.unipd.it/}}}}. Each image has a resolution of $384\times 384$ pixels covering a FOV of $400\times 400 \mu m^2$. The manual annotations of the nerve fibres in these two datasets were traced by an ophthalmologist using the open source software ImageJ.

\textbf{OCTA} dataset is an in-house data collection with 30 retinal OCTA scans. All these scans were acquired using a Heidelberg Spectralis device (Heidelberg, Germany) and all the vessels within the superficial vascular plexus (SVP) were manually traced by a clinical expert using an in-house programme written in Matlab (Mathworks R2018, Natwick) as the ground truth.  In this paper, we use a k-fold (k=5) cross-validation method to divide the training and testing datasets. 

\textbf{OCT RPE} is also an in-house dataset which consists of 36 images of retina vessel shadows projected on the retinal pigment epithelium (RPE) layers of OCT volumes. 
These 3D volumes were captured using a Spectralis OCT system (Heidelberg Engineering GmbH) from 18 healthy volunteers, and have a size of $379 \times 496 \times 384$. The manual annotations of these vessels were labelled by an image analysis expert using the open source software ImageJ. This dataset was originally designed for eliminating retinal vessel shadows in \textit{en face} choroidal OCT.

\begin{figure}[!t]
	\centering{
		\includegraphics[width=0.49\textwidth]{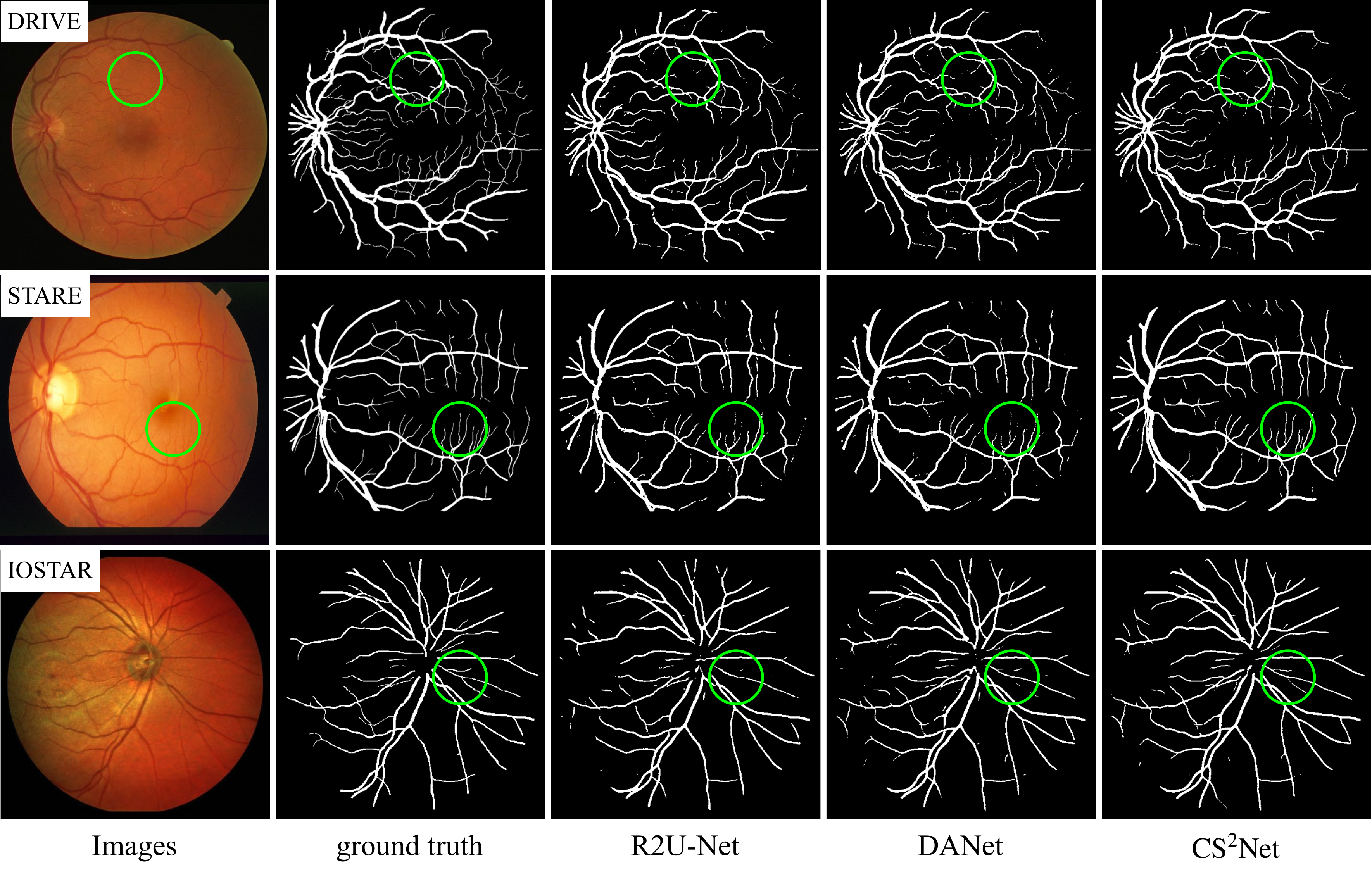}
	}
	\caption{Retinal vessel segmentation results of three randomly selected images from three different datasets by R2U-Net, DANet and our proposed CS$^2$-Net respectively.  }\label{fig_vessel}
\end{figure}

\subsection{Experimental Setup}
The proposed CS$^2$-Net was implemented in the PyTorch library with a dual NVIDIA GPU (GeForce GTX Titan Xp). We use adaptive moment estimation (Adam) as the overall optimizer. The initial learning rate is set to 0.0001 and we use a weight decay of 0.0005 with a poly learning rate policy, where the learning rate is multiplied by $\left( 1- \frac{iter}{max\_iter}\right)^{power}$ with a power of 0.9 and a maximum number of epochs of 100. Due to the limited amount of data, data augmentation is used to improve the performance, which includes random cropping (with a size of $384 \times 384$), contrast enhancement, random rotation ( from -45$^\circ$ to 45$^\circ$ ), random flipping, and mirror flipping about the image centre in the training phase. We do not perform augmentation on the test set. In this paper, we set the batch size to 8 for all the datasets and the proposed method is trained on each imaging modality separately. 

To facilitate the observation and objective evaluation of the proposed method, the  following metrics are adopted, accuracy (\textit{ACC}), sensitivity (\textit{SE}), specificity (\textit{SP}), and Area under the ROC curve (AUC):
	\begin{equation}
	\nonumber
	ACC=\frac{TP+TN}{TP+FP+TN+FN},
	\end{equation}
	\begin{equation}
	\nonumber
	SE=\frac{TP}{TP+FN},\quad SP=\frac{TN}{TN+FP}, 
	\end{equation}
	where TP, FN, TN, and FP denote true positive, false negative, true negative and false positive, respectively. Area under the ROC curve (AUC) reflects the trade-off between sensitivity and specificity, and thus evaluates the quality of our vessel segmentation results more reliably. In addition, we compute the $p$-values of all the evaluation metrics between the proposed method and the compared methods on each dataset for statistical analysis, and $p<$ 0.05 is considered statistically significant.

\subsection{Results}

\begin{figure*}[t]
	\centering{
		\includegraphics[width=0.90\textwidth]{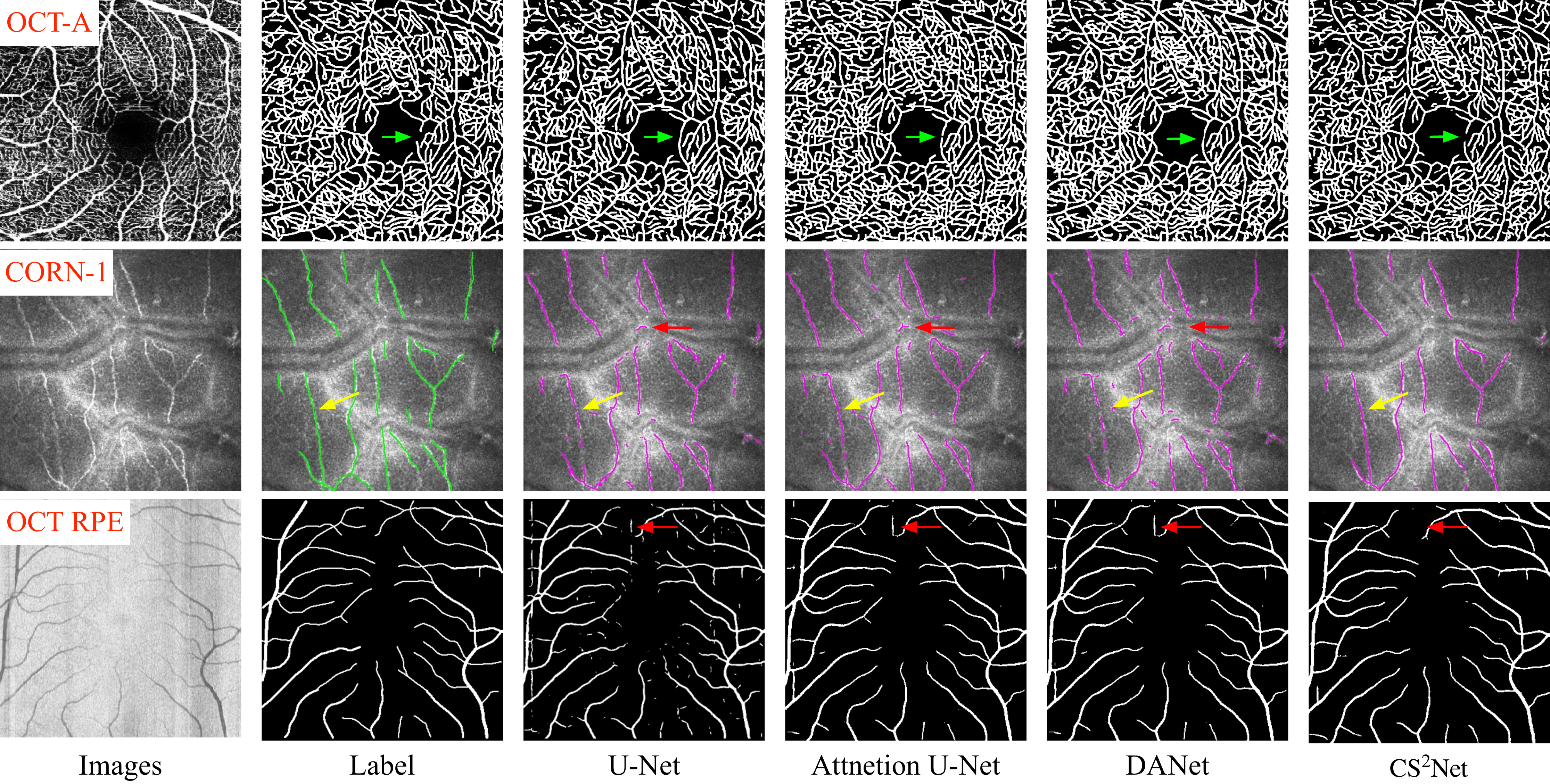}
	}
	\caption{Results of different methods for vessel segmentation of different images in different imaging modalities. From the left to right column: the original images, labels, and segmentation results of U-Net, Attention U-Net, DANet and the proposed CS$^2$-Net, respectively. 
	From the top to bottom row: OCTA, CORN-1 and OCT RPE Layer, respectively. }\label{fig_octa_ccm_oct}
\end{figure*}

\subsubsection{Vessel Segmentation in Color Fundus Image}
To  demonstrate  the  curvilinear  structure  segmentation performance  of  the  proposed  method,  we  first  evaluate it  on  three  public  datasets  (DRIVE, STARE and IOSTAR) that are common and highly recognized in medical imaging. Seven state-of-the-art methods were selected for comparison, which include two conventional filtering-based vessel methods (Combination of Shifted Filter Responses (COSFIRE)~\citep{azzopardi2015trainable} and Weighted Symmetry Filter (WSF)~\citep{ZhaoTMI18}), two specially designed deep learning-based vessel methods (DeepVessel~\citep{Fu2016DeepVesselRV} and Context Encoder Network (CE-Net)~\citep{Gu2019CENetCE}), and three state-of-the-art networks (U-Net~\citep{ronneberger2015u}, Recurrent Residual U-Net (R2U-Net)~\citep{alom2018recurrent}, and Dual Attention Network (DANet)~\citep{fu2018dual}). Note, the results of BCOSFIRE, WSF, and Deep Vessel were quoted from their papers for convenience.

Table~\ref{tab_vessel} shows the segmentation results of different methods on the retinal fundus datasets, where  our proposed CS$^2$-Net outperforms all the competing methods on ACC and AUC scores.
Thus, it can be confirmed that the spatial and channel attention modules are beneficial for retinal vessel detection in colored fundus images. 
Morevoer, Fig.~\ref{fig_vessel} shows the visual comparison between the vessel segmentation results of  R2U-Net~\citep{alom2018recurrent}, DANet~\citep{fu2018dual} and the proposed CS$^2$-Net. We can observe that CS$^2$-Net achieves better performance than R2U-Net and DANet, extracting more vessels in a representative patch (green disc) with multiple scales of vessels in low contrast regions. To better observe the significance of the proposed method and comparison methods in segmenting retinal vessels, we compute the $p$-value for statistical analysis. The results show that the differences between the proposed method and competing methods are significant with all $p$-values $<$0.05.

\begin{table*}[!t]
	\centering
	\caption{Vessel segmentation performances in different metrics of different methods over three retinal fundus datasets. }
	\begin{tabular}{c|l|cccc|c}
		\hline\hline
		\textbf{Datasets}       & \textbf{Methods}    &        \textbf{ACC}         &        \textbf{AUC}         &         \textbf{SE}          &         \textbf{SP}  & $p$-value \\ \hline
		\multirow{8}*{DRIVE} & BCOSFIRE~\citep{azzopardi2015trainable} &       0.9442       &       0.9614       &       0.7655        &       0.9704 & -  \\
		~           & WSF~\citep{ZhaoTMI18}                   &       0.9580       &       0.9750       &       0.7740        &       0.9790      &-  \\
		~           & DeepVessel~\citep{Fu2016DeepVesselRV}   &       0.9533       &       0.9789       &       0.7603        &       0.9776      & - \\
		~           & U-Net~\citep{ronneberger2015u}          &       0.9531       &       0.9601       &       0.7537        &       0.9639      & $<$0.001 \\
		~           & R2U-Net~\citep{alom2018recurrent}       &       0.9556       &       0.9784       &       0.7792        &       0.9813      & 0.019 \\
		~           & CE-Net~\citep{Gu2019CENetCE}            &       0.9545       &       0.9779       & {\bfseries 0.8309 } &       0.9747      & 0.010 \\
		~           & DANet~\citep{fu2018dual}                &       0.9615       &       0.9808       &       0.8075        &       0.9841      & 0.008 \\
		~           & {\bfseries CS$^2$Net}                   &  \textbf{0.9632}   & {\bfseries 0.9825} &       0.8218        & {\bfseries 0.9890}     \\ \hline
		\multirow{8}*{STARE} & BCOSFIRE \citep{azzopardi2015trainable} &       0.9497       &       0.9563       &       0.7716        &       0.9701    & -  \\
		~           & WSF \citep{ZhaoTMI18}                   &       0.9570       &       0.9590       &       0.7880        &       0.9760      & - \\
		~           & DeepVessel \citep{Fu2016DeepVesselRV}   &       0.9609       &       0.9790       &       0.7412        &       0.9701      & - \\
		~           & U-Net \citep{ronneberger2015u}          &       0.9409       &       0.9705       &       0.7675        &       0.9631     & $<$0.001 \\
		~           & R2U-Net \citep{alom2018recurrent}       &       0.9712       &       0.9914       &       0.8298        & {\bfseries 0.9862} & 0.017 \\
		~           & CE-Net \citep{Gu2019CENetCE}            &       0.9583       &       0.9787       &       0.7841        &       0.9725    & 0.009   \\
		~           & DANet \citep{fu2018dual}                &       0.9679       &       0.9781       &       0.7705        &       0.9873    & 0.013  \\
		~           & {\bfseries CS$^2$Net}                   & {\bfseries 0.9752} & {\bfseries 0.9932} & {\bfseries 0.8816}  &       0.9840       \\ \hline
		\multirow{8}*{IOSTAR} & BCOSFIRE \citep{azzopardi2015trainable} & 0.9410 & 0.9550 & 0.7610 & 0.9670  & - \\
		~			 & WSF \citep{ZhaoTMI18}                   &       0.9480       & 0.9600 &       0.7720       &       0.9670   & -     \\
		~           & DeepVessel \citep{Fu2016DeepVesselRV}   &       -       &       -       &       -        &       -   & -   \\
		~           & U-Net \citep{ronneberger2015u}          &       0.9675       &       0.9464       &       0.8044        &       0.9793   & $<$0.001   \\
		~           & R2U-Net \citep{alom2018recurrent}       &       0.9652       &       0.9530       &       0.8042        &  0.9779  &0.014 \\
		~           & CE-Net \citep{Gu2019CENetCE}            &       0.9572      &       0.9658      &       0.8110        &       0.9749     & 0.016  \\
		~           & DANet \citep{fu2018dual}                &       0.9720       &       0.9504       &       0.8298        &       {\bfseries0.9832}   & 0.047    \\
		~           & {\bfseries CS$^2$Net}                   & {\bfseries 0.9722} & {\bfseries 0.9758} & {\bfseries 0.8341}  &       0.9831      \\ \hline
	\end{tabular}
	\label{tab_vessel}
\end{table*}

\subsubsection{Vessel Segmentation in In-house OCTA Images}
To justify that our proposed method can also segment the curvilinear structure on other modal medical images,  we  perform  comparative  experiments  on  our recently released  dataset: In-house  OCTA. We compare the proposed network with five state-of-the-art networks: U-Net~\citep{ronneberger2015u}, Deep ResUNet~\citep{zhang2018road}, U-Net++~\citep{zhou2018unet++},  Attention U-Net~\citep{oktay2018attention}, and DANet \citep{fu2018dual}. The first column of Fig.~\ref{fig_octa_ccm_oct} shows the visual comparison of the vessel segmentation results of different methods on a typical OCTA \textit{en face} image.  Overall, these methods perform well on segmenting significant vessels. Attention U-Net~\citep{oktay2018attention} can detect most significant structures, but it also falsely enhances background features where elongated intensity inhomogeneities are present. U-Net~\citep{ronneberger2015u} mis-detects vessels with small diameters, which leads to a relatively lower sensitivity. In contrast, the proposed CS$^2$-Net adaptively integrates local features with global dependencies and normalization. Hence, it shows superior performance in detecting small vessels, indicated by the green arrow in Fig.~\ref{fig_octa_ccm_oct}, and provides higher sensitivity. These findings are also confirmed by the evaluation measures reported in Table~\ref{tab_octa}, where CS$^2$-Net achieves the highest segmentation performance in terms of all the metrics, since it employs an attention mechanism to build the powerful representation among features. The $p$ values of the proposed method in the pairwise comparison with the the state-of-the-art methods are all less than 0.05, revealing that the proposed method achieves a significant performance improvement over them.

\begin{table*}[t]
	\centering
	\caption{Vessel segmentation performances in different metrics of different methods over our own OCTA dataset.}\label{tab_octa}
	\begin{tabular}{l|cccc|c}
		\hline\hline
		\textbf{Methods}                                   &        \textbf{ACC}         &        \textbf{AUC}         &         \textbf{SE}         &         \textbf{SP}  & $p$-value  \\ \hline
		U-Net~\citep{ronneberger2015u}             &       0.8422       &       0.9108       &       0.7867       &       0.8780    & $<$0.001   \\
		Deep ResUNet~\citep{zhang2018road}         &       0.8659       &       0.9175       &       0.8032       &       0.8863    & $<$0.001    \\
		U-Net++~\citep{zhou2018unet++}              &       0.8965       &       0.9203       &       0.8309       &       0.9101    & 0.017    \\
		Attention U-Net~\citep{oktay2018attention} &       0.9125       &       0.9290       &       0.8274       &       0.9007    & 0.043    \\
		DANet \citep{fu2018dual}                   &       0.8869       &       0.9183       &       0.8427       &       0.8681    & $<$0.001    \\
		{\bfseries CS$^2$Net}                      & {\bfseries 0.9183} & {\bfseries 0.9453} & {\bfseries 0.8631} & {\bfseries 0.9192} \\ \hline
	\end{tabular}
\end{table*}

\begin{table*}[!t]
	\centering
	\caption{Nerve fibre tracing performances in different metrics of different methods over the CORN-1 dataset (mean $\pm$ standard deviation).}\label{tab_nerve}
	\begin{tabular}{l|cc|c}
		\hline\hline
		\textbf{Methods} & \textbf{SE} $\uparrow$ & \textbf{FDR} $\downarrow$ & $p$-value\\
		\hline
		U-Net~\citep{ronneberger2015u} &  0.7757$\pm$0.0144 &  0.3961$\pm$0.0208 & $<$0.001 \\
		Deep ResUNet~\citep{zhang2018road} &  0.8038$\pm$0.0140 &  0.2911$\pm$0.0214  & 0.003\\
		U-Net++~\citep{zhou2018unet++} & 0.8274$\pm$0.0127 &  0.2715$\pm$0.0118 & 0.015 \\
		Attention U-Net~\citep{oktay2018attention} & 0.8166$\pm$0.0131 & 0.2761$\pm$0.0120 &  0.013 \\
		DANet \citep{fu2018dual} & 0.8012$\pm$0.0043 & 0.3850$\pm$0.0011 & $<$0.001 \\
		{\bfseries CS$^2$Net} & \textbf{0.8398$\pm$0.0098} & \textbf{0.2556$\pm$0.0028}   \\
		\hline
	\end{tabular}
\end{table*}

\begin{figure*}[t]
\centering
\includegraphics[width=\textwidth]{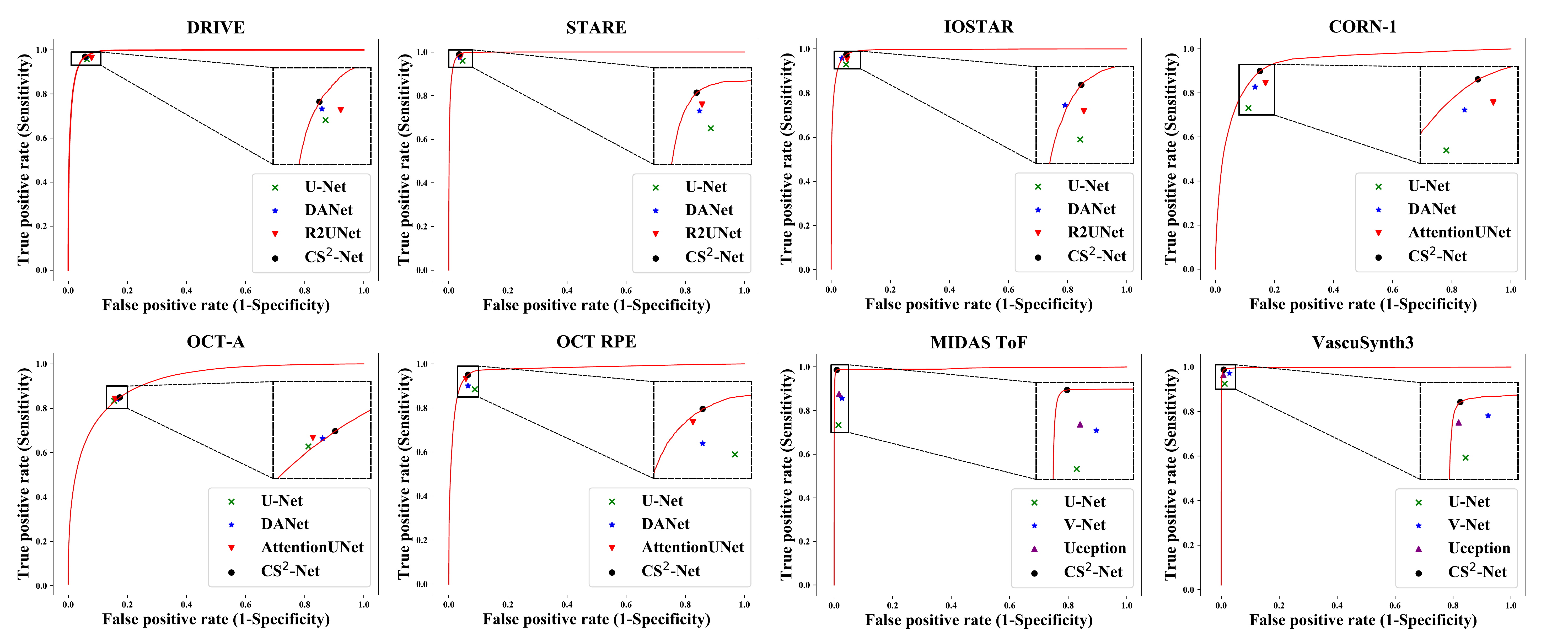}
\caption{ROC curves of our proposed CS$^2$-Net for curvilinear structure segmentation over different datasets: DRIVE, STARE, IOSTAR, CORN-1, OCT-A, OCT RPE, MIDAS and VascuSynth3 datasets, compared with those of the state-of-the-art methods at particular TP and FP rates.}
\label{fig-roc}
\end{figure*}

\subsubsection{Corneal Nerve Fiber Tracing in the CORN-1 Images}
We further evaluate the performance of our CS$^2$-Net for corneal nerve fiber tracing  on the CORN-1 dataset that we have  published. For validation, we compute the sensitivity and  $\textit{false discovery rate}$ (FDR)~\citep{guimaraes2016fast} between the predicted centerlines of the nerve fibres and groundtruth.  FDR is defined as the fraction of the total pixels incorrectly detected as nerve segments over the total pixels of the traced nerves in the ground truth. As is customary in the evaluation process~\citep{guimaraes2016fast}, if any pixel on the extracted pixel-wide curves is within the three-pixel tolerance region around the manually traced nerves, it is a true positive. 

Similar to the evaluation on the OCTA images, we again employed U-Net~\citep{ronneberger2015u}, Deep ResUNet~\citep{zhang2018road}, U-Net++~\citep{zhou2018unet++}, Attention U-Net~\citep{oktay2018attention}, and DANet~\citep{fu2018dual}  as the baselines for comparison. The second row of Fig.  \ref{fig_octa_ccm_oct} illustrates a sample image from the CORN-1 dataset. Although all methods present visually appealing results, both U-Net, Attention U-Net and DANet falsely detect parts of the K-structures~\citep{Yokogawa} (indicated by the red arrows) as nerve fibres, because they share similar morphological characteristics. In contrast, our CS$^2$-Net ensures continuous fibre tracing (indicated by the yellow arrows). Table~\ref{tab_nerve} shows the performances of different methods for fibre tracing  on the CORN-1 dataset. The challenge of corneal nerve fibre tracing is to preserve the continuity of the fibers. As a basic network, U-Net~\citep{ronneberger2015u} performs worse than the other methods.  Deep ResUNet~\citep{zhang2018road} and DANet~\citep{fu2018dual} obtain similar results in SE. Our method achieves the best tracing performance in terms of either the SE or FDR. In addition, the $p$-value of the proposed method is less than 0.015, which shows that there is a significant difference in performance between the proposed method and the state-of-the-art methods.


\begin{table*}[t]
	\centering
	\caption{Vessel segmentation performances in different metrics of different methods over our OCT RPE Layer dataset.}
	\begin{tabular}{l|cccc|c}
		\hline\hline
		\textbf{Methods}                                   &       \textbf{ACC}       &       \textbf{AUC}       &       \textbf{SE}        &       \textbf{SP}  & $p$-value \\ \hline
		U-Net~\citep{ronneberger2015u}             &     0.9550      &     0.9370      &     0.7875      &     0.9807   & $<$0.001 \\
		Deep ResUNet~\citep{zhang2018road}         &     0.9601      &     0.9591      &     0.8071      &     0.9838   & 0.019     \\
		U-Net++~\citep{zhou2018unet++}              &     0.9654      &     0.9578      &     0.8138      &     0.9822   & 0.028     \\
		Attention U-Net~\citep{oktay2018attention} &     0.9664      &     0.9584      &     0.8142      &     0.9813   & 0.043     \\
		DANet \citep{fu2018dual}                   &     0.9686      &     0.9667      &     0.7849      &     \textbf{0.9850}   & 0.027   \\
		\textbf{CS$^2$Net}                         & \textbf{0.9693} & \textbf{0.9686} & \textbf{0.8296} & 0.9840     \\ \hline
	\end{tabular}
	\label{tab-oct}
\end{table*}

\subsubsection{Vessel Segmentation in the OCT RPE Layers}
The proposed method is also validated on another different modal dataset for curviliear structure segmnetation: OCT RPE Layers. The vascular projections in the RPE layers are not true blood vessels. However, they can be considered important features to assist artefact removal on the choroid. We use the same metrics as those for the color fundus vascular segmentation to evaluate the performance of the RPE vascular projection segmentation methods. Similarity, we use U-Net~\citep{ronneberger2015u}, Deep ResUNet~\citep{zhang2018road}, U-Net++~\citep{zhou2018unet++}, Attention U-Net~\citep{oktay2018attention}, and DANet~\citep{fu2018dual} to make comparisons with the proposed CS$^2$-Net. Metric scores are shown in Table~\ref{tab-oct}, which demonstrates the superior vascular projection performance of our model, and there are also significant differences in performance among the comparison methods, indicated by $p=0.041$.  The last row of Fig.  \ref{fig_octa_ccm_oct} shows a randomly selected RPE image, in which the proposed method clearly demonstrates more resistance to the interference caused by the capturing device. The proposed method extracts tiny blood vessels more effectively than either U-Net~\citep{ronneberger2015u} or DANet~\citep{fu2018dual}, and it does not produce over-segmentation, as indicated by the red arrows.

In Fig.~\ref{fig-roc}, we show the ROC curves of our proposed CS$^2$-Net over different datasets for the segmentation of curvilinear structures:  DRIVE, STARE, IOSTAR, CORN-1, OCT-A, and OCT RPE, compared with those of the state-of-the-art methods at particular TP and FP rates for the sake of readability. It can be seen from the local enlarged view of Fig.~\ref{fig-roc} that the proposed method outperforms on the whole state-of-the-art methods for curvilinear structure segmentation, despite the variation of structure, contrast and imaging noise from one imaging modality to another.

\section{Experimental Results over 3D Volumes}

\subsection{Materials}

To further demonstrate the broad applicability of the proposed method for the segmentation of 3D vasculatures in different modalities, we evaluate our method over 3D volumes from three publicly-accessible datasets: one brain MRA dataset (i.e., MIDAS) and two  synthetic datasets (i.e., Synthetic, and VascuSynth).

\textbf{MIDAS}{\footnote{\url{http://hdl.handle.net/1926/594}}} is a publicly available MRA dataset. This dataset contains 50 MRA volumes acquired from 25 male and 25 female healthy volunteers, aged from 18 to 60+ years. Images  were captured using a 3T MRI scanner under standardized protocols, with a voxel size of $0.5\times 0.5\times 0.8$ mm$^3$. These were reconstructed as a $448 \times 448 \times 128$ matrix. Manual annotations of Circles of Willis (CoW) were  provided by Prof. Alejandro Frangi  from the University of Leeds,  where 3D vasculatures were generated by  tracing the centerlines of the vessels, and the vessel surfaces were  extracted using the geodesic active contour method~\citep{Bogunovic2011}.

\textbf{Synthetic}{\footnote{\url{https://github.com/giesekow/deepvesselnet/wiki/Datasets}}} was  originally generated  using the method proposed in~\citep{schneider2012tissue}, and includes 136 volumes of size $325 \times 304 \times 600$ with their corresponding labels for vessel segmentation, centerlines and bifurcation detection. 

\textbf{VascuSynth}{\footnote{\url{http://vascusynth.cs.sfu.ca/Data.html}}} aims to provide an abundance of 3D images for the automated analysis of tree-like structures, which includes vessel segmentation and detection of bifurcation points using the VascuSynth Software \citep{jassi2011vascusynth}. It simulates volumetric images (a size of $100 \times 100 \times 100$ voxels) of vascular trees and generates the corresponding ground truth for segmentation, bifurcation locations, branch properties, and tree hierarchy.

\subsection{Experimental Setup}
The proposed 3D CS$^2$-Net was implemented in the PyTorch framework with a dual NVIDIA GPU (Titan Xp). Adam serves as the optimizer for all comparative experiments. We adopt a poly learning strategy with an initial learning rate of 0.0001 and a weight decay of 0.0005. Due to the different sizes of 3D volumes, we have different crop sizes for different datasets, the experimental details can be found in the following subsections. Besides, we normalized the volume including training and test data and set the maximum training iteration to 200.

To better evaluate the binary segmentation performance of the proposed 3D CS$^2$-Net, we follow \citep{zhao2018automatic} and adopt the following metrics:  true positive rate (TPR),  false negative rate (FNR), and false positive rate (FPR).
	To demonstrate our model can learn more vascular features from sparsely labelled annotations and has  better discrimination ability for non-vascular patterns, we introduce two new metrics, over-segmentation rate (OR) and under-segmentation rate (UR), to evaluate the model:
	\begin{equation}
	\nonumber
	OR=\frac{O_s}{R_s+O_s}, \quad UR=\frac{U_s}{R_s+O_s},
	\end{equation}
	where $R_s$ denotes all the voxels inside the ground truth, $O_s$ denotes the voxels inside the predicted volume but not inside the ground truth, and $U_s$ indicates the voxels inside the ground truth but not in the predicted volume. According to the definition, it can be seen that $OR\in [0,1]$ and $UR \in [0,1]$. The lower the values of these metrics, the better the performance of the method.

\begin{table*}[t]
	\centering
	{\small
	\caption{Vessel segmentation performances in different metrics of different methods over the MRA dataset.}
	\begin{tabular}{l|ccccc|c} 	\hline\hline
		\textbf{Methods}                &  \textbf{TPR} $\uparrow$  & \textbf{FNR} $\downarrow$ & \textbf{FPR} $\downarrow$ &  \textbf{OR} $\downarrow$  & \textbf{UR} $\downarrow$ & $p$-value \\ \hline
		MVEF~\citep{frangi1998multiscale} &     0.9143      &      0.0424       &      0.0648       &            -             &            -       & -    \\
		IUWF~\citep{bankhead2012fast}    &     0.9387     &      0.0402     &      0.0602       &            -             &            -     & -        \\
		QFMS~\citep{lathen2010blood}      &     0.9512      &      0.0383      &      0.0591      &            -             &            -   & -         \\
		WSF~\citep{zhao2018automatic}    &     0.9678     &      0.0342      &      0.0562     &            -             &            -   & -         \\
		3D U-Net~\citep{unet3d}           &     0.9521      &      0.0479     &      0.0053     &     {0.0833}      &     0.0393   & $<$0.001   \\
		V-Net~\citep{milletari2016v} & 0.9616  & 0.0483  & 0.0009  & 0.1043  & 0.0352 & $<$0.001  \\
		Uception~\citep{uception} & 0.9567  & 0.0433   & 0.0006   & 0.2005   & 0.0318 & $<$0.001 \\
		\textbf{CS$^2$Net}               & \textbf{0.9683} & \textbf{0.0285}  & \textbf{0.0004}  & \textbf{0.0801} & \textbf{0.0291}  \\ \hline
	\end{tabular}
	\label{tab_brainvessel}
	}
\end{table*}

\subsection{Brain Vessel Segmentation in MRA Volumes}

In this section, we  evaluate the proposed curvilinear structure segmentation method on the cerebral MRA images. Since the manual annotations in the MIDAS dataset are sparse, i.e., many vascular voxels do not have labels. Here, the metrics such as Dice coefficient (DC) and Intersection over Union (IoU) are not appropriate to validate its performance, since there are significantly more non-vascular voxels than the vascular ones. 
We perform a center crop of the raw data along the axial plane with a size of $224\times 224 \times 64$. While the original labels are triangular polygon surfaces which cannot be directly used as input for 3D convolutions, we employ the open source medical image processing toolkit \textit{The Visualization Toolkit} (VTK)\footnote{\url{https://vtk.org/}} to voxelize these surfaces. These operations significantly reduce the size of the volumes, which allows us to set a larger batch size of 2 in this paper. To better justify the vessel shape and structure extraction performance of the proposed 3D CS$^2$-Net in the real-world scenarios, we finally evaluate it over the MRA images. Under this setting, we compare the proposed method with six state-of-the-art methods: 3D Multi-scale Vessel Enhancement Filtering (MVEF)~\citep{frangi1998multiscale}, 3D Isotropic Undecimated Wavelet Filtering (IUWF)~\citep{bankhead2012fast}, 3D Quadrature Filters across Multiple Scales (QFMS)~\citep{lathen2010blood}, 3D Weighted Symmetry Filter (WSF)~\citep{zhao2018automatic}, V-Net~\citep{milletari2016v}, 3D U-Net~\citep{unet3d}, and Uception~\citep{uception}. To validate the cerebrovascular segmentation performance, we compute the TPR, FNR, FPR, OR and UR between the predicted volume and ground truth. We obtained the results of TPR, FNR and FPR from~\citep{frangi1998multiscale, bankhead2012fast} and~\citep{ZhaoTMI18} and put them into Table~\ref{tab_brainvessel}. We also use the OpenSource code of 3D U-Net to train the model and carefully fine-tune it to reach optimality. The evaluation metrics are computed according to the predicted results. All the results are shown in Table~\ref{tab_brainvessel}. We can see that the proposed method achieves better performance in the 3D cerebrovascular segmentation task, surpassing all the other methods in terms of TPR, FNR and FPR.

The proposed method reduces the FPR of the other six methods by 0.0644, 0.0598, 0.0587, 0.0548, 0.0049 and 0.0005, respectively. This means that the proposed 3D CS$^2$-Net is better at distinguishing cerebrovasculatures from the complex background artefacts in MRA images. We can conclude from the UR metric in Table~\ref{tab_brainvessel} that the predicted cerebrovasculatures of both the 3D U-Net and the proposed method achieve the highest similarity with the ground truth, since the under-segmentation rates are only 0.0393 and 0.0291, respectively. However, the proposed 3D CS$^2$-Net achieves  a  lower  under-segmentation  rate  by 0.0102. Based on the TPR metric, we can see that the proposed method can segment the cerebrovasculatures with the highest segmentation rate (up to 0.9706). The larger OR achieved by 3D U-Net indicates that more unlabelled cerebrovascular vessels are segmented as vascular ones. Overall, the proposed method shows better reliability for the segmentation of cerebral blood vessels in terms of TPR, UR and OR, Moreover, compared with the selected methods, the proposed method gains a $p$-value of less than 0.001, which shows that the proposed method is significantly better than other methods in segmentation performance. 

\begin{figure}[t]
	\centering
	\includegraphics[width=0.48\textwidth]{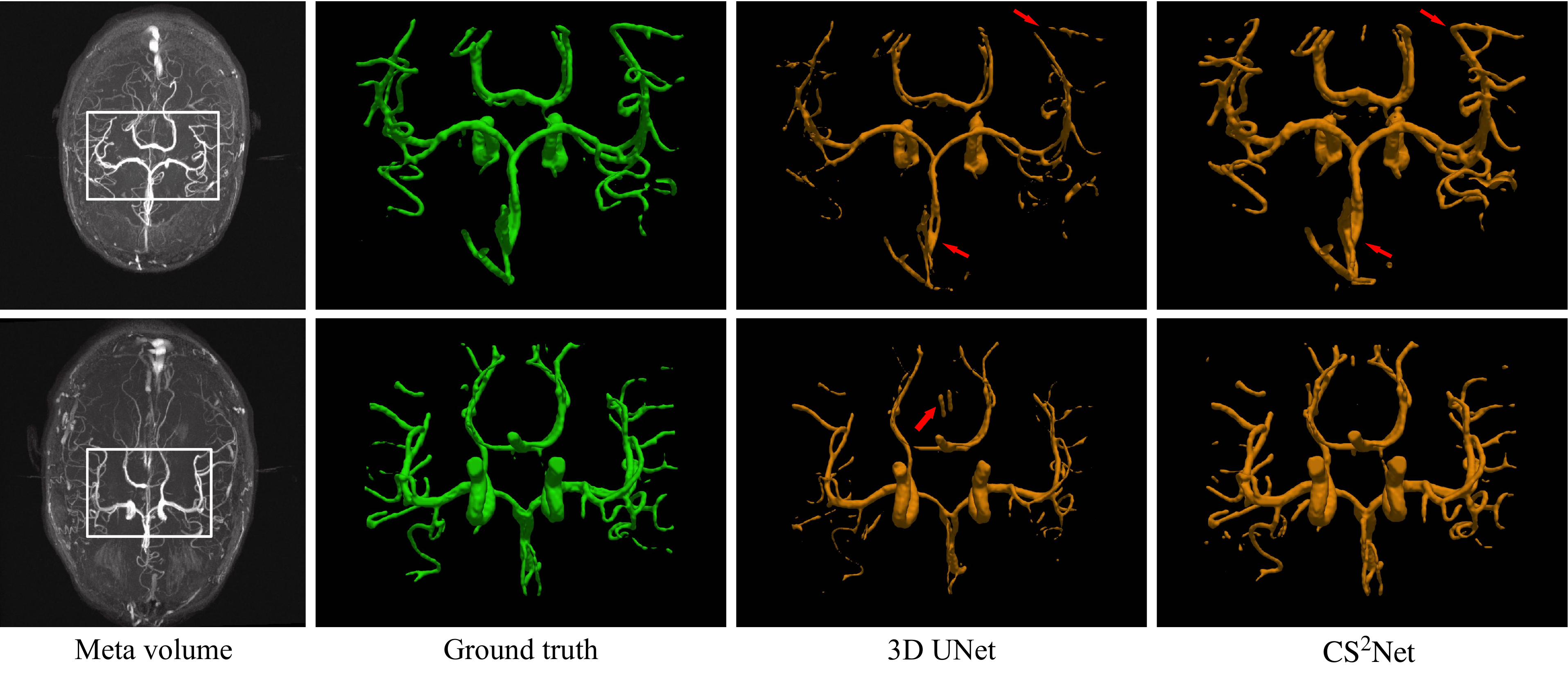}
	\caption{3D renderings of curvilinear structure segmentation results of an image in the MRA dataset. 
	From the left to right column: a MIP view of a sample MRA image, the segmentation of ground truth, the 3D U-Net and  the proposed CS$^2$-Net respectively.}
	\label{synthetic_brain}
\end{figure}

Fig.~\ref{synthetic_brain} shows the segmentation results of 3D U-Net and the proposed 3D CS$^2$-Net on one image in the MRA dataset. Our 3D CS$^2$-Net presents better vascular extraction performance than 3D U-Net, especially for  tiny vessels, as indicated by  the red arrows in Fig.~\ref{synthetic_brain}. On the other hand, it can be observed from Fig.~\ref{synthetic_brain} that the cerebral vessels segmented by 3D U-Net are thinner than the ground truth, while those segmented by the proposed 3D CS$^2$-Net are more similar. Therefore, 3D U-Net tends to under-segment or miss cerebral vessels, which can be verified from the OR and UR metrics in Table~\ref{tab_brainvessel} respectively.

\begin{table*}[t]
	\centering
	\caption{Vessel segmentation results in different metrics of different methods over different 3D datasets.}
	\begin{tabular}{c|l|cccc|c}
		\hline\hline
		\textbf{Datasets}           & \textbf{Methods}                         & \textbf{TPR} $\uparrow$ & \textbf{FNR} $\downarrow$ & \textbf{FPR} $\downarrow$ & \textbf{DC} $\uparrow$ & $p$-value \\ \hline
		\multirow{3}*{Synthetic}   & 3D U-Net~\citep{unet3d}           &     0.9965      &       0.0035       &       0.0001       &     0.9106  & 0.025   \\
		~ & V-Net~\citep{milletari2016v} & 0.9949 & 0.0051 & 0.0001 & 0.9237 & 0.027 \\
		~              & Uception~\citep{uception}        &  0.9984        &    0.0026       &   0.0003   &   0.9785  & 0.032 \\
		~              & \textbf{CS$^2$Net}               & \textbf{0.9986} &  \textbf{0.0014}   &  \textbf{0.0000}   & \textbf{0.9913}   \\ \hline
		\multirow{6}*{VascuSynth-1} & ITM~\citep{cetin2012vessel}      &     0.9423      &       0.0577       &       0.0471       &     0.9406   & -   \\
		~              & CBS~\citep{cheng2015accurate}    &     0.9529      &       0.0471       &       0.0563       &     0.9489    & - \\
		~              & WSF~\citep{zhao2018automatic}    &     0.9678      &       0.0342       &       0.0562       &     0.9601    & -  \\
		~              & 3D U-Net~\citep{unet3d}           &     0.9704      &       0.0096       &       0.0007       &     0.9552    & $<$0.001  \\
		~              & V-Net~\citep{milletari2016v}     &     0.9763      &      0.0088       &       0.0003       &     0.9594     & 0.011 \\
		~              & Uception~\citep{uception}        &     0.9800      &    0.0071       &   0.0004   &   0.9426   & 0.008 \\
		~              & \textbf{CS$^2$Net}               & \textbf{0.9841} &  \textbf{0.0068}   &  \textbf{0.0001}   & \textbf{0.9637}  \\ \hline
		\multirow{6}*{VascuSynth-2} & ITM~\citep{cetin2012vessel}      &     0.9423      &       0.0577       &       0.0471       &     0.9406   & -   \\
		~              & CBS~\citep{cheng2015accurate}    &     0.9529      &       0.0471       &       0.0563       &     0.9489  & -    \\
		~              & WSF~\citep{zhao2018automatic}    &     0.9603      &  \textbf{0.0451}   &       0.0526       &     0.9543   & -   \\
		~              & 3D U-Net~\citep{unet3d}           &     0.9602      &       0.0502       &       0.0013       &     0.9587   & 0.009   \\
		~              & V-Net~\citep{milletari2016v}     &     0.9605      &       0.0503       &       0.0011       &     0.9584   & 0.015      \\
		~              & Uception~\citep{uception}        &  0.9607        &    0.0510       &   0.0009   &   0.9468   & 0.013 \\
		~              & \textbf{CS$^2$Net}               & \textbf{0.9611} &       0.0494       &  \textbf{0.0004}   & \textbf{0.9593}   \\ \hline
		\multirow{3}*{VascuSynth-3} & 3D U-Net~\citep{unet3d}           &     0.9338      &       0.0661       &       0.0024       &     0.9112  & $<$0.001   \\
		~              & V-Net~\citep{milletari2016v}     &     0.9365      &       0.0598       &       0.0027       &     0.9037    & $<$0.001  \\
		~              & Uception~\citep{uception}        &  0.9413        &    0.1033       &   0.0066   &   0.9157  & $<$0.001 \\
		~              & \textbf{CS$^2$Net}               & \textbf{0.9484} &  \textbf{0.0416}   &  \textbf{0.0005}   & \textbf{0.9256}   \\ \hline
	\end{tabular}
	\label{tab_vascusynth}
\end{table*}

\begin{figure}[t]
	\centering
	\includegraphics[width=0.48\textwidth]{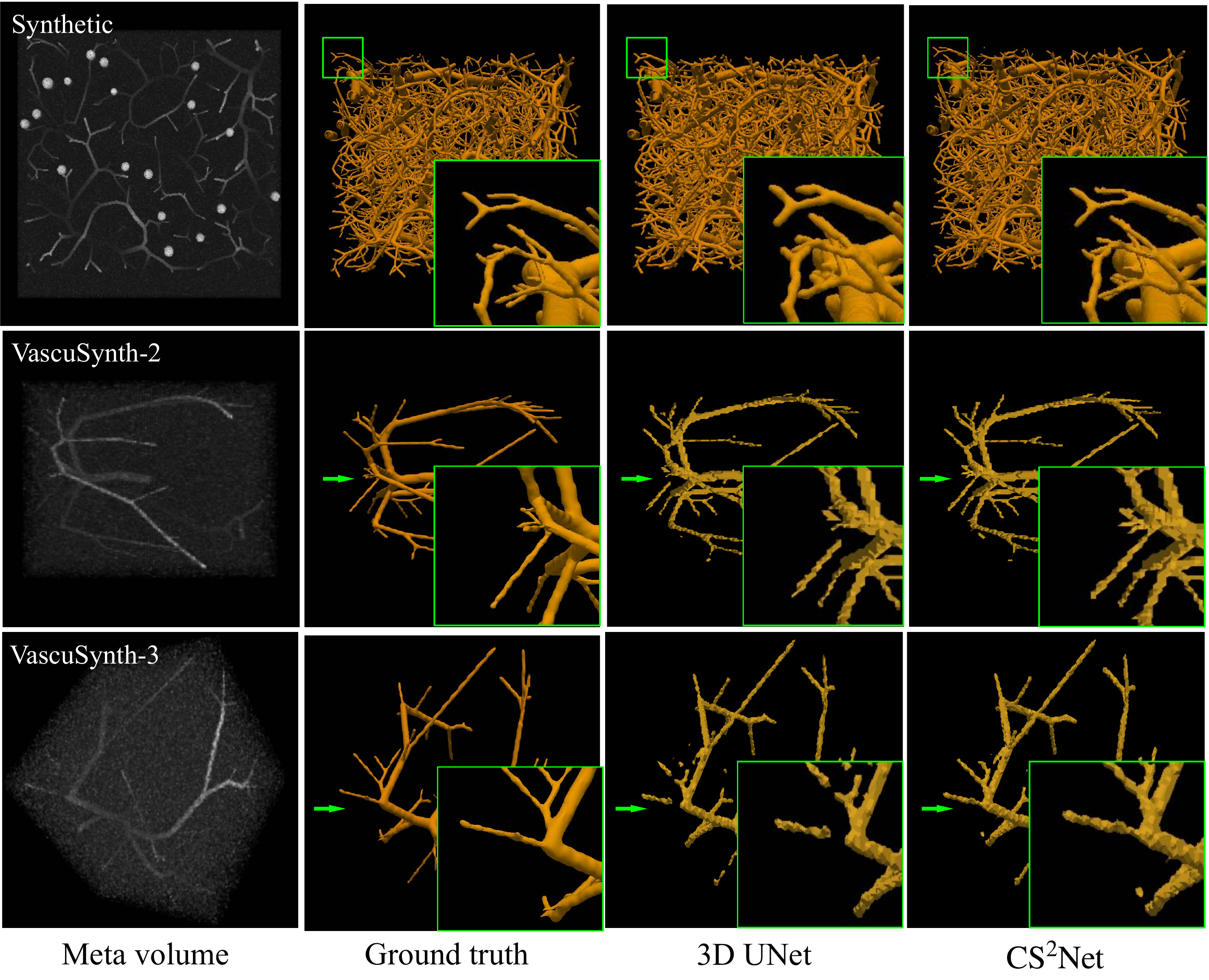}
	\caption{3D renderings of curvilinear structure segmentation results of different methods over \textbf{Synthetic} and \textbf{VascuSynth}. The first column shows volumes with the different levels of noise  ($\sigma^2=20$ for Synthetic, $\sigma^2=60$ for VascuSynth-2 and $\sigma^2=100$ for VascuSynth-3). 
	Segmentation results of different methods in the second to right column: ground truth, 3D U-Net and the proposed CS$^2$-Net, respectively. The green boxes in different rows show an enlarged view of the local segmentation results.}
	\label{fig-vascusynth}
\end{figure}

\subsection{Vessel Segmentation in Synthetic Data}

To further demonstrate the advantage of the proposed 3D CS$^2$-Net, we also report its segmentation performance over two synthetic datasets:  Synthetic and VascuSynth. For both datasets, we apply k-fold ($k=4$) cross-validation  to divide the training and testing datasets, i.e. 5 randomly selected volumes serve as the testing set and the remaining ones are used to train the model.
In addition,  Gaussian noise with three standard variances $\sigma^2$ is added to the  VascuSynth dataset to mimic imaging artefacts, to investigate how the proposed method behaves in detecting curvilinear structures in noise-corrupted data. In the remainder of this paper, we refer to these noise-corrupted versions as: VascuSynth-1 ($\sigma^2 = 20$), VascuSynth-2 ($\sigma^2 = 60$), and VascuSynth-3 ($\sigma^2 = 100$).
An additional random crop operation with a size of $128\times 128 \times 128$ is adopted to reduce the training cubes. We set the batch size to 6 in this part.

We first evaluate our proposed method and compare it with
the state-of-the-art ones: Intensity-based Tensor Model (ITM)~\citep{cetin2012vessel}, Constrained B-Snake (CBS) ~\citep{cheng2015accurate}, Weighted Symmetry Filtering (WSF)~\citep{zhao2018automatic},  3D U-Net~\citep{unet3d}, V-Net~\citep{milletari2016v} and Uception~\citep{uception} on the Synthetic dataset. The results are shown in Table~\ref{tab_vascusynth}. We observe that the proposed method successfully segments 3D curvilinear structures with competitive performance and outperforms the 3D U-Net in terms of TPR, FNR, FPR, and particularly DC (up by 0.0807). Fig.~\ref{fig-vascusynth} further demonstrates the segmentation performance of both methods. Compared with the 3D U-Net~\citep{unet3d}, the proposed method shows better discrimination ability at boundaries, which can be verified from the enlarged view (green box) of the first row in Fig.  \ref{fig-vascusynth}. The synthetic vessels segmented by the 3D U-Net are thicker than those of the proposed method. This proves that the proposed method can achieve better edge discrimination ability through the proposed CSAM. It can also be analysed from $p=0.032<0.05$ that the proposed method performs significantly better than either the 3D U-Net, the V-Net or Uception.

We further make comparisons between some of state-of-the-art methods and the proposed method. All the performances are evaluated based on TPR, FNR, FPR and DC. We follow \citep{zhao2018automatic} and add Gaussian noise with a standard variance of $\sigma^2=20$ to generate the VascuSynth-1 dataset. To thorough verify the proposed model, we also follow \citep{zhao2018automatic} to compare the proposed method with other state-of-the-art models (Intensity based Tensor Model (ITM)~\citep{cetin2012vessel},  Constrained B-Snake (CBS) ~\citep{cheng2015accurate}, WSF~\citep{zhao2018automatic}), and deep learning-based model (3D U-Net~\citep{unet3d}, V-Net~\citep{milletari2016v}), and Uception~\citep{uception} on VascuSynth with different standard variances of $\sigma^2=20$ and $\sigma^2=60$ respectively. 
These experimental results are shown in Table \ref{tab_vascusynth}. As can be seen,  the proposed CS$^2$-Net outperforms the state-of-the-art methods in the segmentation of 3D curvilinear structures. 
Fig.~\ref{fig-vascusynth} illustrates the 3D segmentation results of two sample 3D images by the proposed method and 3D U-Net. As indicated by the green arrow and representative patches,  we can observe that the 3D U-Net detects discontinuous vessels and misses the small ones in the middle left of the figure. In sharp contrast, the proposed CS$^2$-Net detects all the vessels more thoroughly, even though they vary in thickness, length, and local contrast with the background. For VascuSynth-1 and VascuSynth-2, the proposed method has better performance in segmenting the 3D curvilinear structure than other methods, confirmed by $p=0.011$ and $p=0.015$ respectively.

Since 3D U-Net, V-Net, Uception and the proposed network are deep learning-based methods, we apply a higher-level noise of $\sigma^2=100$ on the volume data to further confirm their performance for 3D curvilinear structure segmentation. Quantitative results  are shown in Table \ref{tab_vascusynth}. 
The results of different methods for the curvilinear structure segmentation of a randomly selected image in the VascuSynth-3 dataset are presented in the last row of Fig.~\ref{fig-vascusynth}. From the table and the detailed view in the green box of the figure, we can conclude that the proposed method has still detected vessels more completely compared with the other methods. This is because the attention model in the proposed method evaluates the expression capability of the features globally over the whole images and normalises them in the feature space and are thus more robust to the local noise and variation in size of the vessels. Compared with 3D U-Net, V-Net, and Uception as the noise level increases, the performance of the proposed method increases significantly, which can be concluded from the change of the $p$-value from $0.01<p<0.05$ ($\sigma^2=20$ and $\sigma^2=60$) to $p<0.001$ ($\sigma^2=100$).

\section{Discussions}

The proposed CS$^2$-Net utilizes spatial and channel attention modules to capture the structural information of the tree-like objects in the horizontal and vertical directions, respectively. In this work, we carefully designed a network focusing on the extraction of the curvilinear structures in medical images. Compared with natural images, medical imagery contains unique features, such as simpler semantics and unitary patterns. Therefore, we first construct a network backbone based on the encoder-decoder framework. More importantly, we introduce a $1\times 3$ and a $3 \times 1$ convolutional kernel to capture more boundary features to assist the segmentation of curvilinear structures. DANet~\citep{fu2018dual} uses a pre-trained model to extract features, and up-samples the attention features in the last layer of the model, and this is the architectural difference between the proposed method and DANet. Second, we introduce batch normalization and ReLU activations after the convolutional layers in the spatial attention module to ensure that the mean and variance of the input distribution are fixed within a specific range, reducing the internal covariate shift in the network, and mitigating the gradient disappearance to a certain extent. Third, since 3D volume contains rich depth information not included in 2D medical images, many critical lesions can be better observed through different layers in the 3D volume. Here, we extend the 2D attention mechanism to 3D to enhance the network's ability to aggregate depth information across different image slices. Therefore, we design a 3D volume segmentation network, and introduce $1\times 1 \times 3$, $1 \times 3 \times 1$ and $3\times 1 \times 1$ convolution kernels in the 3D attention module with batch normalization and ReLU activetions. This network is thus more suitable for 3D medical data analysis. Besides, we evaluate the proposed method on a variety of medical datasets in different modalities, and the evaluation results also confirmed that our proposed method is effective for segmenting curvilinear structures.

\begin{figure}[!t]
	\centering
	\includegraphics[width=0.48\textwidth]{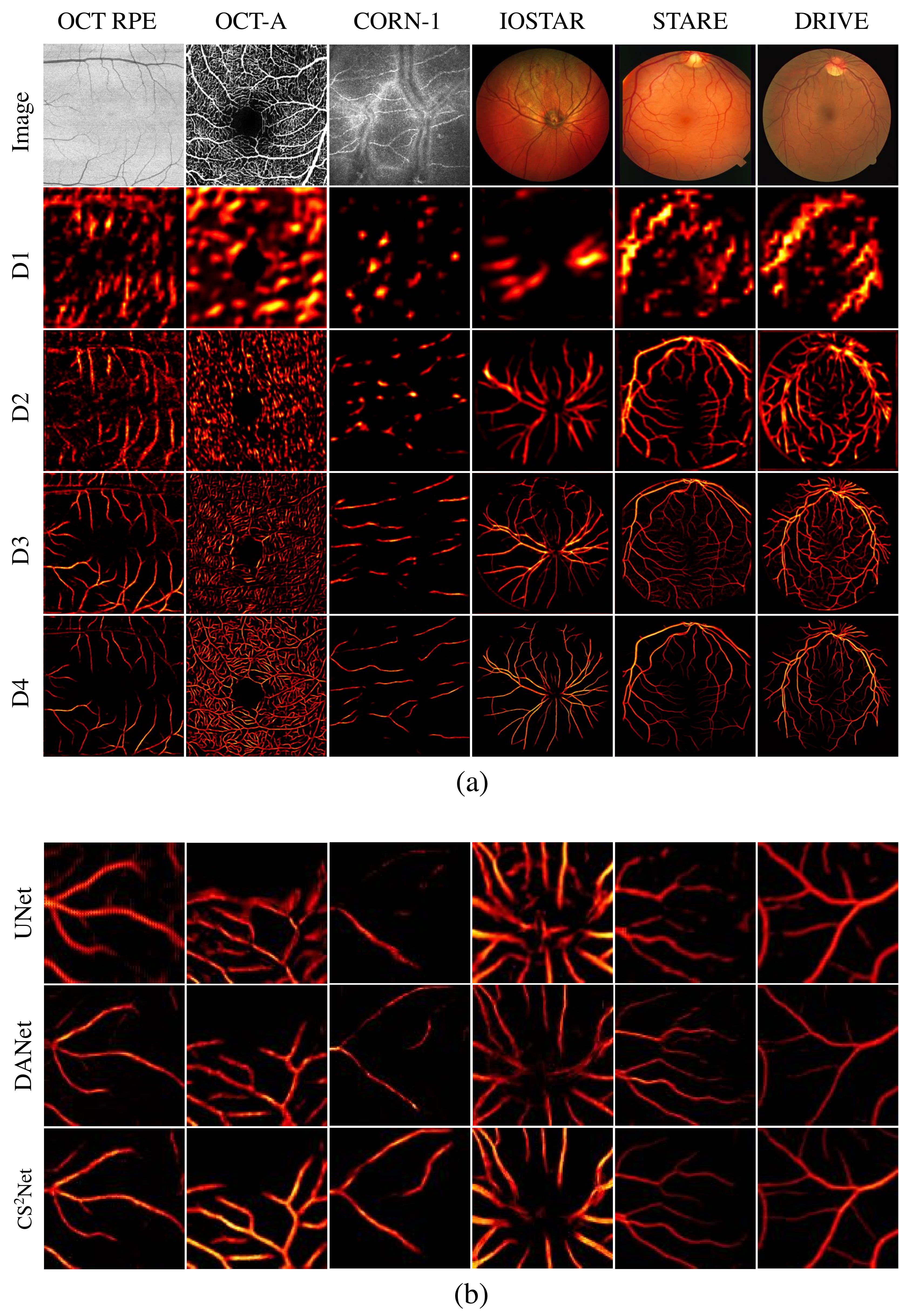}
	\caption{Attention maps of different methods in the intermediate layers of the decoding parts. 
	(a) the attention maps of the proposed CS$^2$-Net in different decoding layers on different datasdets:  DRIVE, STARE, IOSTAR, CORN-1, OCT-A, and OCT RPE datasets, respectively. D1 $\sim$ D4 display the attention maps representing the incremental refinement in curvilinear structure segmentation; (b) the enlarged local intermediate attention maps of different methods: U-Net, DANet, and CS$^2$-Net. }
	\label{fig-heatmap}
\end{figure}

\begin{table}[t]
	\centering
	\small
	\caption{TPR, FNR, FPR, Over-segmentation Rate (OR) and Under-segmentation Rate (UR) of the proposed method with a combining of different components for the curvilinear structure segmentation of the 3D images in the MIDAS ToF MRA dataset. }
	\begin{tabular}{l|ccccc}
		\hline
		\hline
		\textbf{Methods} & \textbf{TPR}  $\uparrow$ & \textbf{FNR} $\downarrow$ & \textbf{FPR} $\downarrow$ & \textbf{OR} $\downarrow$ & \textbf{UR} $\downarrow$\\
		\hline
		Backbone & 0.9517 & 0.0493 & 0.0103 & 0.0877 & 0.0808  \\
		Backbone+CAB & 0.9663 & 0.0310 & 0.0024 & 0.1082 & 0.0341   \\
		Backbone+SAB & 0.9565 & 0.0413 & 0.0018 & 0.1147 & 0.0532 \\
		\textbf{CS$^2$Net} & \textbf{0.9706} & \textbf{0.0285} & \textbf{0.0004} & \textbf{0.1027} & \textbf{0.0296} \\
		\hline 
	\end{tabular}
	\label{tab-ablation}
\end{table}

To better support improved segmentation results, we further visualize intermediate attention maps in our proposed CS$^2$-Net over different datasets, as shown in Fig.~\ref{fig-heatmap}(a). By analysing and comparing the blood vessels and nerve fibres in the attention maps from $D1$ to $D4$, we note that the proposed model can focus on curvilinear structures during training. The curvilinear structures gradually become brighter and smoother. In low-level attention maps, the highlighted areas basically are distributed around the curvilinear structure regions, which reflect that the CAB module focuses more classified information on the curvilinear structure. On the other hand, the highlighted areas at different spatial locations also confirm that the SAB module can enhance the ability of the proposed network to capture long-range dependencies of curvilinear structure. In addition, we present several sets of attention maps of the proposed method and two selected state-of-the-art methods (DANet and U-Net) in the same intermediate layer in Fig.~\ref{fig-heatmap}(b) to gain intuition and verify the influence of the attention modules. Overall, it can be observed from the comparison of each column that the proposed CS$^2$-Net has stronger response than both DANet and U-Net in terms of curvilinear structure information aggregation. Here, the proposed CS$^2$-Net is more responsive to vessels and nerve fibres than DANet, which can be clearly seen from the brighter highlights of curvilinear structures in Fig.~\ref{fig-heatmap}(b). By comparing the attention maps of CS$^2$-Net and U-Net, it can be seen clearly that  the proposed CS$^2$-Net is more powerful in suppressing the background interference than U-Net.

To demonstrate the effectiveness of the 3D CSAM in  CS$^2$-Net,  we carry out an ablation study over the MIDAS dataset. First, we test the backbone of our network, e.g., Deep ResUNet~\citep{zhang2018road}, without the CSAM. For fair comparison, we retrain the backbone network under the same hyperparameter settings as the proposed method and use the same metrics to perform its evaluation. Second, we perform a further ablation study by removing the CAB but retaining the SAB inside the original CS$^2$-Net. For the final ablation study, we remove the SAB in the CS$^2$-Net but retain the CAB to form the final set of the ablation study. All computed metrics are shown in Table \ref{tab-ablation}. The results reveal that the proposed CSAM can effectively extract the features of curvilinear structures. 
The backbone performance (Backbone) is slightly improved copared to the CAB only (Backbone+CAB) from 0.9517 to 0.9663 in TPR and from 0.0493 to 0.0310 in FNR. However, much better performance is achieved with the SAB only (Backbone+SAB).  This is because the CAB normalises the features in the feature space for simple binary classification (vascular and non-vascular) tasks. At the same time, the SAB enhances the features over the whole image and thus increases the contrasts of different objects at different locations. Thus, the CSAM integrates the advantages of both the CAB and SAB to make the model better at producing inter-class discrimination and intra-class responses, and thus obtains the best performance with a TPR of 0.9706  and an FNR of 0.0285.

\begin{figure}[!t]
	\centering
	\includegraphics[width=0.48\textwidth]{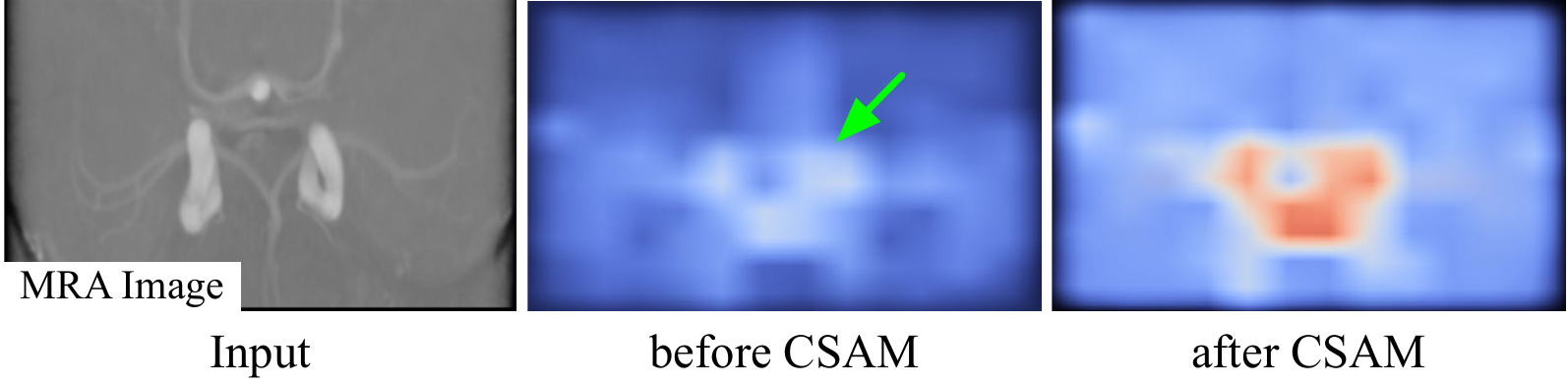}
	\caption{The output of the proposed CSAM on a randomly selected image from the MIDAS dataset. From the left to right: the original volume, the predicted probabilities of voxles as curvilinear structure before and after applying the proposed CSAM, respectively.}
	\label{fig-before_after}
\end{figure}

Besides numerical verification, we also obtain and visualise the outputs before and after CSAM for different image modalities through heat maps. We applied up-sampling and sigmoid operations on the outputs to resize the feature maps to the corresponding image size and normalize the outputs to $[0, 1]$, respectively. The visualization results are shown in Fig.~\ref{fig-before_after}. As indicated by the green arrow, the voxels with the probabilities as being curvilinear structure by our proposed CSAM are better clustered and indicative 
than those without it. These results show that the CSAM can extract and aggregate the edge features of the curvilinear structures and enhance the network's ability to distinguish between tubular and non-tubular patterns. 

\section{Conclusion and Future Works}
Curvilinear structure segmentation is a fundamental step in automated diagnosis of many diseases, and it remains a challenging medical image analysis problem despite considerable research efforts. In this paper, we developed a new curvilinear structure segmentation network, named CS$^2$-Net, which applies to both 2D images and 3D volumes. Our CS$^2$-Net improves the inter-class discrimination and intra-class aggregation abilities, by applying a self-attention mechanism to high-level features in the channel and spatial dimensions \citep{fu2018dual}. The experimental results over 9 datasets across 6 imaging modalities have demonstrated that the proposed method can improve segmentation results. Our results confirm its great potential as a powerful image analysis method application in computer-aided diagnosis from medical imaging, and in automated biological image interpretation.

Over the past few years there has been an increasing number of AI models proposed and published. The lack of evaluation of their usefulness across different images in different applications shows down their adoption in real applications. Our work makes a first step towards more extensive evaluation of AI models to demonstrate their effectiveness and applicability across several applications.

Although this paper highlights the potential and applicability of our proposed CS$^2$-Net method for general curvilinear structure segmentation, there remain several areas of improvement and future research. Diseased cells with similar features to curvilinear structures can lead our model to over segment these curvilinear structures. Additional information could help addressing this limitation like, for instance, the inclusion of local neighbourhood and continuity constraints. Three-dimensional volumetric segmentation consumes considerable GPU resources, which increases the computational demands of model training. The architecture could potentially be simplified without compromising accuracy, and hence reducing the complexity of the proposed method.

\section*{Acknowledgments}
This work was supported by grants from the Zhejiang Provincial Natural Science Foundation (LZ19F010001 and LQ20F030002), the Key Research and Development Program of Zhejiang Province (2020C03036), National Science Foundation Program of China (61906181), and Ningbo \lq\lq 2025 S\&T Megaprojects" (2019B10033 and 2019B10061). AFF is supported by the RAEng Chair in Emerging Technologies (INSILEX Pnogramme CiET1819/19), European Union’s Horizon 2020 Research and Innovation Programme (InSilc SC1-PM-16-2017-777119), Cancer Research UK funding Leeds Radiotherapy Research Centre of Excellence (CRUK RadNet C19942/A28832), and the Pengcheng Visiting Scholars Award at Shenzhen University from Shenzhen Ministry of Education.
\bibliographystyle{model2-names.bst}\biboptions{authoryear}
\bibliography{reference-MedIA}

\end{document}